\documentclass[a4paper,11pt]{article}
\pdfoutput=1
\usepackage[dvipsnames]{xcolor}
\usepackage[T1]{fontenc} 
\usepackage[toc,page]{appendix}
\usepackage{jcappub} 
\usepackage{circledsteps}
\usepackage{natbib}
\setcitestyle{square,comma,numbers,sort&compress}
\usepackage{enumerate}
\usepackage{mathrsfs}
\usepackage{tabularx,booktabs}
\usepackage[left = 1.0in, top=1.0in,right=1.0in,bottom=1.0in,a4paper]{geometry}
\usepackage[caption=false]{subfig}
\usepackage[compat=1.1.0]{tikz-feynman} 
\usepackage{cancel}
\def\mc#1{\mathcal#1}
\usepackage[section]{placeins}
\usepackage{cleveref}

\usepackage{simpler-wick}
\newcolumntype{Y}{>{\centering\arraybackslash}X}

\definecolor{darkgreen}{rgb}{0,0.5,0}

\definecolor{myRED}{rgb}{0.8, 0.25, 0.33}

\title{NANOGrav signal from axion inflation}

\author[a]{Xuce Niu,}
\author[b]{and Moinul Hossain Rahat}

\affiliation[a]{Institute for Fundamental Theory, Department of Physics,
University of Florida,\\Gainesville, FL 32611, USA }
\affiliation[b]{School of Physics \& Astronomy, University of Southampton, Southampton SO17 1BJ, UK }
\emailAdd{xuce.niu@ufl.edu}
\emailAdd{M.H.Rahat@soton.ac.uk}

\abstract{ 
Several pulsar timing arrays including NANOGrav, EPTA, PPTA, and CPTA have recently reported the observation of a stochastic background of 
gravitational wave spectrum in the nano-Hz frequencies. An inflationary interpretation of this observation is challenging from various aspects. We report that such a signal can arise from the Chern-Simons coupling in axion inflation, where a pseudoscalar inflaton couples to a (massive) $U(1)$ gauge field, leading to efficient production of a transverse gauge mode. Such tachyonic particle production during inflation exponentially enhances the primordial perturbations and leads to a unique parity-violating gravitational wave spectrum, that remains flat near the CMB scales but becomes blue-tilted at smaller scales. We identify the parameter space consistent with various cosmological constraints and show that the resultant gravitational wave signals can provide extra contribution on top of the standard astrophysical contribution from inspiraling supermassive black hole binaries towards explaining the 
observed excess at NANOGrav. The parity-violating nature of the signal can be probed in future interferometers, distinguishing it from most other new physics signals attempting to explain the NANOGrav result.

}

\begin{document}
\maketitle
\flushbottom

\section{Introduction}
The first detection of gravitational wave (GW) signals from two colliding black holes by the LIGO and VIRGO  collaborations \cite{LIGOScientific:2016aoc} heralded a new era in observational astronomy. Since then, the gravitational wave interferometers have detected signals from various astrophysical sources. However, the possibility of detection of a stochastic background of gravitational wave has remained elusive until very recently, when several pulsar timing arrays (PTA) including EPTA \cite{EPTA:2023sfo, EPTA:2023akd, EPTA:2023fyk, EPTA:2023gyr, EPTA:2023xxk, EPTA:2023xiy}, PPTA \cite{Reardon:2023gzh, Reardon:2023zen, Zic:2023gta}, IPTA \cite{Antoniadis:2022pcn}, CPTA \cite{Xu:2023wog}, and finally NANOGrav \cite{NANOGrav:2023gor, NANOGrav:2023hvm, NANOGrav:2023hfp,NANOGrav:2023icp, NANOGrav:2023ctt, NANOGrav:2023hde, NANOGrav:2023tcn} announced the observation of excess red common-spectrum signals in the nano-Hz regime, with inter-pulsar correlations following the Hellings-Downs pattern \cite{Hellings:1983fr}, pointing to a stochastic gravitational wave background (SGWB) origin. The result from various PTAs are consistent and  similar, we therefore focus on the results obtained from NANOGrav 15 yr (NG15) dataset.

The source of this SGWB can be various, ranging from supermassive black hole binaries \cite{NANOGrav:2023hfp} to new physics \cite{NANOGrav:2023hvm}, such as cosmic inflation \cite{Guth:1980zm,Lyth:1998xn, Kinney:2003xf, Baumann:2009ds}, scalar-induced gravitational waves \cite{Domenech:2021ztg, Yuan:2021qgz}, first order phase transitions \cite{Athron:2023xlk}, topological defects like cosmic strings  or domain walls \cite{ Kibble:1976sj,Vilenkin:1981bx,Vilenkin:2000jqa}, with varying degrees of likelihood \cite{NANOGrav:2023hvm}. See refs. \cite{Buchmueller:2023nll, Broadhurst:2023tus, Yang:2023aak, Yang:2023qlf, Eichhorn:2023gat, Huang:2023chx, Wang:2023ost, Gouttenoire:2023ftk, DiBari:2023upq, Cai:2023dls, Inomata:2023zup, Lazarides:2023ksx, Depta:2023qst, Blasi:2023sej, Franciolini:2023wjm, Shen:2023pan, Zu:2023olm, Lambiase:2023pxd, Han:2023olf, Guo:2023hyp, Wang:2023len, Ellis:2023tsl, Vagnozzi:2023lwo, Fujikura:2023lkn, Kitajima:2023cek, Li:2023yaj, Franciolini:2023pbf, Megias:2023kiy, Ellis:2023dgf, Bai:2023cqj, Ghoshal:2023fhh, Deng:2023btv, Rini:2023eqy, Athron:2023mer, Addazi:2023jvg, Oikonomou:2023qfz, Kitajima:2023vre, King:2023cgv, Zhao:2022kvz, Vagnozzi:2020gtf, Benetti:2021uea, Chen:2019xse, Kobakhidze:2017mru, Arunasalam:2017ajm} for some recent examples.

In this paper we focus on a possible origin of this signal from an axion inflation model \cite{Niu:2022quw,Niu:2022fki, Adshead:2015pva, Adshead:2016iae, Adshead:2018doq, Adshead:2019lbr, Adshead:2019igv, Freese:1990rb, Silverstein:2008sg, McAllister:2008hb, Kim:2004rp, Berg:2009tg, Dimopoulos:2005ac, Pajer:2013fsa}, in which the pseudoscalar inflaton $\phi$ with an approximate shift symmetry \cite{Freese:1990rb} has a Chern-Simons coupling $\phi F \tilde{F}$ to a $U(1)$ gauge field, where $F$ is the field strength of the gauge field and $\tilde{F}$ is its dual. This coupling enables a tachyonic production of a transverse mode of the gauge field \cite{Anber:2006xt,Anber:2009ua, Cook:2011hg, Barnaby:2010vf, Barnaby:2011qe, Barnaby:2011vw, Meerburg:2012id, Anber:2012du, Linde:2012bt, Cheng:2015oqa, Garcia-Bellido:2016dkw, Domcke:2016bkh, Domcke:2016mbx, Peloso:2016gqs, Domcke:2018eki, Cuissa:2018oiw}, which 
enhances the primordial scalar and tensor perturbations. On one hand, it introduces significant non-Gaussianity in the scalar perturbations, on the other hand, it yields a characteristic {parity-violating} GW spectrum that remains flat at low CMB frequencies but rises at smaller scales probed by the gravitational wave interferometers. Intriguingly, the backreaction of the gauge fields on the inflationary dynamics tames the growth of the GW signal, helping it evade stringent constraints from cosmological considerations and non-observation of SGWB at LIGO scales \cite{KAGRA:2021kbb}. The degree of backreaction, which determines at what frequency the signal starts to rise, depends on the inflationary potential. Here we consider the 
T-model \cite{Carrasco:2015rva}
as a specific example currently favored by Planck 2018 data \cite{Planck:2018jri}, and show that the GW signal becomes blue-tilted at small enough frequencies to reach the amplitude range reported by the PTAs, above the expected contribution from inspiraling supermassive black hole binaries (SMBHBs), while remaining consistent with various cosmological constraints. 

The production of massive gauge fields during inflation has at least two more important aspects. First, it generates a {parity-violating} GW spectrum coming from two polarizations of the graviton, a unique feature that distinguishes it from most other sources of SGWB. Such parity-violating nature can be probed in future interferometers \cite{Orlando:2020oko, Badger:2021enh}. Second, it has interesting implications for ``cosmological collider'' physics \cite{Chen:2009zp,  Chen:2009we, Baumann:2011nk, Arkani-Hamed:2015bza, Chen:2016nrs, Lee:2016vti, Meerburg:2016zdz, Chen:2016uwp, Chen:2016hrz, An:2017hlx, Kumar:2017ecc, Chen:2018xck, Wu:2018lmx, Li:2019ves, Lu:2019tjj, Hook:2019zxa, Hook:2019vcn, Kumar:2019ebj, Wang:2019gbi, Wang:2020uic, Wang:2020ioa, Aoki:2020zbj, Maru:2021ezc, Lu:2021gso, Wang:2021qez, Tong:2021wai, Cui:2021iie, Pinol:2021aun, Tong:2022cdz, Reece:2022soh, 
Jazayeri:2022kjy, Pimentel:2022fsc, Chen:2022vzh, Qin:2022lva, Maru:2022bhr, Aoki:2023tjm, Chen:2023txq}, where the mass and spin of a particle produced during inflation leave imprints on the oscillatory three-point correlation function of the scalar perturbations in the ``squeezed limit''. The fact that the same process generates gravitational wave signals potentially observed at PTAs offers a remarkable opportunity to probe cosmological collider physics in a scale well beyond the CMB/LSS scales.

The paper is organized as follows. In section 2, we review the phenomenology of gauge field production in axion inflation and discuss relevant constraints on the model. We discuss the evolution of the inflationary variables from CMB scales to smaller scales in the context of the 
T-model potential
in section 3. In section 4, we discuss the gravitational wave signals in this model and show that the results provide extra contribution on top of SMBHBs towards explaining the NANOGrav 15-yr signal at face value. We conclude in section 5.

\section{Phenomenology of gauge boson production}\label{sec:axioninflation}
We contemplate a single field slow-roll inflation scenario, where the inflaton $\phi$ is an axion-like pseudoscalar with an approximate shift symmetry 
$\phi \to \phi +  {\rm const}$. 
Due to the shift symmetry, the inflaton  couples to a $U(1)$ 
gauge field $A_\mu$ via a Chern-Simons term, $- \phi F \tilde{F} / (4 \Lambda)$, where $F_{\mu\nu} \equiv \partial_\mu A_\nu -\partial_\nu A_\mu$ is the field strength of the gauge field, 
and $\Tilde{F}^{\mu\nu} \equiv \frac{1}{2} \frac{\epsilon^{\mu\nu\alpha\beta}}{\sqrt{-g}}F_{\alpha\beta}$ is its dual. The action can be written as 
\begin{align}
    S = \int d^4x \sqrt{-g} \left[   - \frac{1}{2}\partial_\mu \phi \partial^\mu \phi - V(\phi)
        - \frac{1}{4}F^{\mu\nu}F_{\mu\nu}+\frac{1}{2}m_A^2 A^\mu A_\mu - 
         \frac{1}{4\Lambda}\phi \Tilde{F}^{\mu\nu}F_{\mu\nu} \right], \label{action}
\end{align}
where $\epsilon^{0123} = +1$ is a tensor antisymmetric in any two indices.
We assume a quasi-de Sitter space with a slowly varying Hubble rate $H$, with the scale factor given by $a(t) = e^{Ht}$. The metric can be expressed as    
\begin{align}
    ds^2 \equiv g_{\mu\nu}dx^\mu dx^\nu 
      = dt^2 - a^2(t) \delta_{ij} dx^i dx^j 
   = a^2(\tau) (d\tau^2 - \delta_{ij} dx^i dx^j),
\end{align}
where $t$ and $\tau$ are the physical and conformal time, respectively.

We assume that the inflaton field has a homogeneous background, on top of which we consider an inhomogeneous perturbation, 
$\phi (t, {\bf x}) = \phi_0(t) + \delta \phi(t, {\bf x} ).$ This allows us to study the background dynamics and the correlation functions of the perturbations separately.

The dynamics of the background inflaton field can be expressed in terms of the following coupled equations 
\begin{gather}
    \ddot{\phi}_0 + 3 H\dot{\phi}_0+\frac{dV}{d\phi_0} = \frac{1}{\Lambda} \langle\mathbf{E}\cdot\mathbf{B}\rangle, \label{backr1}\\
    3 H^2 M_{\rm Pl}^2 - \frac{1}{2}\dot{\phi}_0^2 - V =  \frac{1}{2}\left\langle \mathbf{E}^2 + \mathbf{B}^2+\frac{m_A^2}{a^2} \mathbf{A}^2 \right\rangle, \label{backr2} 
\end{gather}
where the physical `electric' and `magnetic' fields corresponding to the gauge field are given by $\mathbf{E} = - \mathbf{A'}/{a^2}$ and $ \mathbf{B} =  (\boldsymbol{\nabla} \times \mathbf{A})/{a^2}$. The terms on the right hand side act as sources of backreaction of the gauge fields on the inflationary dynamics, and can have a significant impact on observables at and beyond the CMB scale. Backreaction effects can no longer be neglected when the source terms on the right hand side are comparable to the terms on the left hand side.

For the general case of massive gauge bosons, we employ the constraint $\partial_\mu (\sqrt{-g}A^\mu) = 0$ derived from the equation of motion and 
decompose the gauge field in the helicity basis $\lambda = \pm , 0$, 
\begin{align}
    \mathbf{A}(\tau, \mathbf{x}) = \sum_{\lambda = \pm,0} \int \frac{d^3 k}{(2\pi)^3} 
      \left[ \boldsymbol{\epsilon}_\lambda (\mathbf{k}) a_\lambda (\mathbf{k}) A_\lambda (\tau, k) 
         e^{i \mathbf{k}\cdot \mathbf{x}} + \text{h.c.} \right] \label{gaugedecomp}
\end{align}
where the polarization vector $ \boldsymbol{\epsilon}_\lambda (\mathbf{k})$ and the annihilation 
operator $a_\lambda (\mathbf{k})$ obey the usual commutation and orthonormality relations. $A_\lambda (\tau, k)$ is the mode function, where the longitudinal mode and two transverse modes are denoted by $\lambda =0$ and $\lambda = \pm$, respectively.  

The dominant vector field production is governed by the field equations of the transverse modes,
\begin{align}
   \partial_\tau^2{{A}_\pm} ( \tau, k) + \left(k^2 + {a(\tau)^2} m_A^2 \pm \frac{2k\xi}{\tau}\right) {{A}_\pm} ( \tau, k)  &= 0,
   \label{eq:Aeq}
\end{align}
where $\xi \equiv {\dot{\phi}_0}/{(2\Lambda H)}$ and we have used $a \approx -1/(H\tau)$ during inflation. Taking $\dot{\phi}_0 > 0$ without loss of generality, only the $A_+$ mode experiences tachyonic instability and is dominantly produced. 
We assume the usual Bunch-Davies initial condition and treat the inflaton's rolling speed $\dot{\phi}_0$ to be a constant during inflation. The solution of the mode function can then be written in terms of the Whittaker-W function,
\begin{equation}
    {A}_\pm (\tau, k ) =  
     \frac{1} {\sqrt{2k}} e^{\pm\pi \xi/2} W_{\mp i\xi, i\mu}(2ik\tau), \label{Whittaker}
\end{equation}
where $\mu \equiv \sqrt{(m_A/H)^2 - 1/4}$. The $A_-$ mode is exponentially suppressed, and the longitudinal mode is unaffected by this production mechanism, but can otherwise be produced purely from gravitational effects \cite{Graham:2015rva, Ema:2019yrd, Ahmed:2020fhc, Kolb:2020fwh, Fong:2022cmq}.  
The mode function solution for the massless case is derived by setting $m_A = 0$ in \cref{Whittaker}. For a particular $k$ mode, the energy density of the gauge mode function $A_+$ receives an exponential enhancement for $\xi > m_A/H$ when $-k\tau \sim O(1)$, which overcomes the Boltzmann suppression and enhances particle production, 
so that the physical observables impacted by the gauge boson production have an approximately exponential dependence on $\xi - m_A/H$ \cite{Niu:2022quw}. We treat $\xi$ and $m_A/H$ as free parameters of the model that are later constrained from various phenomenological considerations.

We now turn to the effect of gauge field production on primordial scalar and tensor perturbations.
The inflaton perturbations follow the equation of motion \cite{Niu:2022quw}
\begin{align}
     &\delta\ddot{ \phi} + 3\beta H \delta\dot{ \phi} - \left(\frac{1}{a^2}\nabla^2  - \frac{d^2 V}{d\phi^2}\right)\delta \phi   = \frac{1}{\Lambda} \left(\mathbf{E} \cdot \mathbf{B} - \langle \mathbf{E} \cdot \mathbf{B}\rangle\right), \label{EOMscalarperturb2}
\end{align}
where $\beta \equiv 1- 2\pi\xi {\langle \mathbf{E} \cdot \mathbf{B}\rangle}/{(3\Lambda H\dot{\phi}_0)}$. From the inflaton perturbations, the curvature perturbation on uniform density hypersurfaces is defined as 
\begin{align}
\zeta (\tau, \mathbf{x}) \equiv -\frac{{H}}{\dot{\phi}_0}\ \delta \phi(\tau, \mathbf{x}),
\end{align}

To calculate the tensor perturbation, we write the perturbed metric in terms of the tensor perturbation $h_{ij}$ using the scalar-vector-tensor decomposition as
\begin{align}
    ds^2 = a^2(\tau) \left[ d\tau^2 - (\delta_{ij} + h_{ij})dx^i dx^j \right],
    \label{eq:hij_definition}
\end{align}
where $h_{ij}$ is transverse ($\partial_i h_{ij} = 0$) and traceless ($h_{ii} = 0$). The equation of motion of $h_{ij}$ is given by \cite{Weinberg:2008zzc}
\begin{align}
    h_{ij}''-\nabla^2 h_{ij} + 2 \mc H h_{ij}'=\frac{2}{M_{\rm Pl}^2} T_{ij}^{TT}, \label{eomh}
\end{align}
where $M_{\rm Pl} \simeq 2.4\times 10^{18}$ GeV is the reduced Planck mass, and $T_{ij}^{TT}$ is the transverse and traceless part of the stress-energy tensor. We decompose the tensor perturbation into two helicity modes
\begin{align}
    h_{ij}(\tau, \mathbf{p}) = \sum_{\lambda = \pm} \epsilon^\lambda_i(\mathbf{p})\epsilon^\lambda_j(\mathbf{p})\left(a_\lambda(\mathbf{p}) h_{p}^\lambda(\tau) + a^\dagger_\lambda(\mathbf{-p}) h_{p}^{\lambda*}(\tau)\right) \equiv
     \sum_{\lambda = \pm} \epsilon^\lambda_i(\mathbf{p})\epsilon^\lambda_j(\mathbf{p}) h^\lambda(\tau, {\bf p}).
\end{align}

The correlation functions of the curvature and tensor perturbations are calculated at $\tau_0 = 0$ after the end of inflation. For details of the calculation using the in-in formalism \cite{Weinberg:2005vy}, we refer the interested readers to refs. \cite{Niu:2022quw, Niu:2022fki}.
Here we focus on the phenomenological observables derived from the two and three-point correlation functions of the primordial perturbations. The scalar power spectrum is proportional to the two-point correlation function of the curvature perturbation and can be written as
$P_\zeta = P_\zeta^{[\phi]} + P_\zeta^{[A]}$,
where $P_\zeta^{[\phi]} \equiv \left(\frac{H}{\dot{\phi}_0}\right)^2 \left(\frac{H}{2\pi} \right)^2$ comes from the usual vacuum fluctuations, and 
\begin{align}
   P_\zeta^{[A]} \equiv \frac{2 k^3}{(2\pi)^2}\langle\zeta_{\mathbf{k_1}}(\tau_0)\zeta_{\mathbf{k_2}}(\tau_0)\rangle_{(1)}^\prime  
   \label{PzetaA}
\end{align}
comes from the impact of the gauge field production. Here ${}^\prime$ denotes that the $\delta$-function $(2\pi)^3\delta^{(3)}(\mathbf{k_1} + \mathbf{k_2})$ is stripped off. The amplitude of the scalar power spectrum at the CMB scale is measured to be
$P_\zeta \simeq 2.4 \times 10^{-9}$ \cite{WMAP:2010qai, Bunn:1996py},
which accounts for the contribution of the vacuum modes as well as the extra degrees of freedom (gauge modes in this case). Conservatively taking the gauge field's contribution to be subdominant at the CMB, we can ignore $P_\zeta^{[A]}$ and fix $P_\zeta^{[\phi]} = 2.4 \times 10^{-9}$. This assumption would be valid as long as $P_\zeta^{[A]} \ll P_\zeta^{[\phi]}= 2.4 \times 10^{-9}$.  
\begin{figure}
    \centering
    \includegraphics[width=0.65\textwidth]{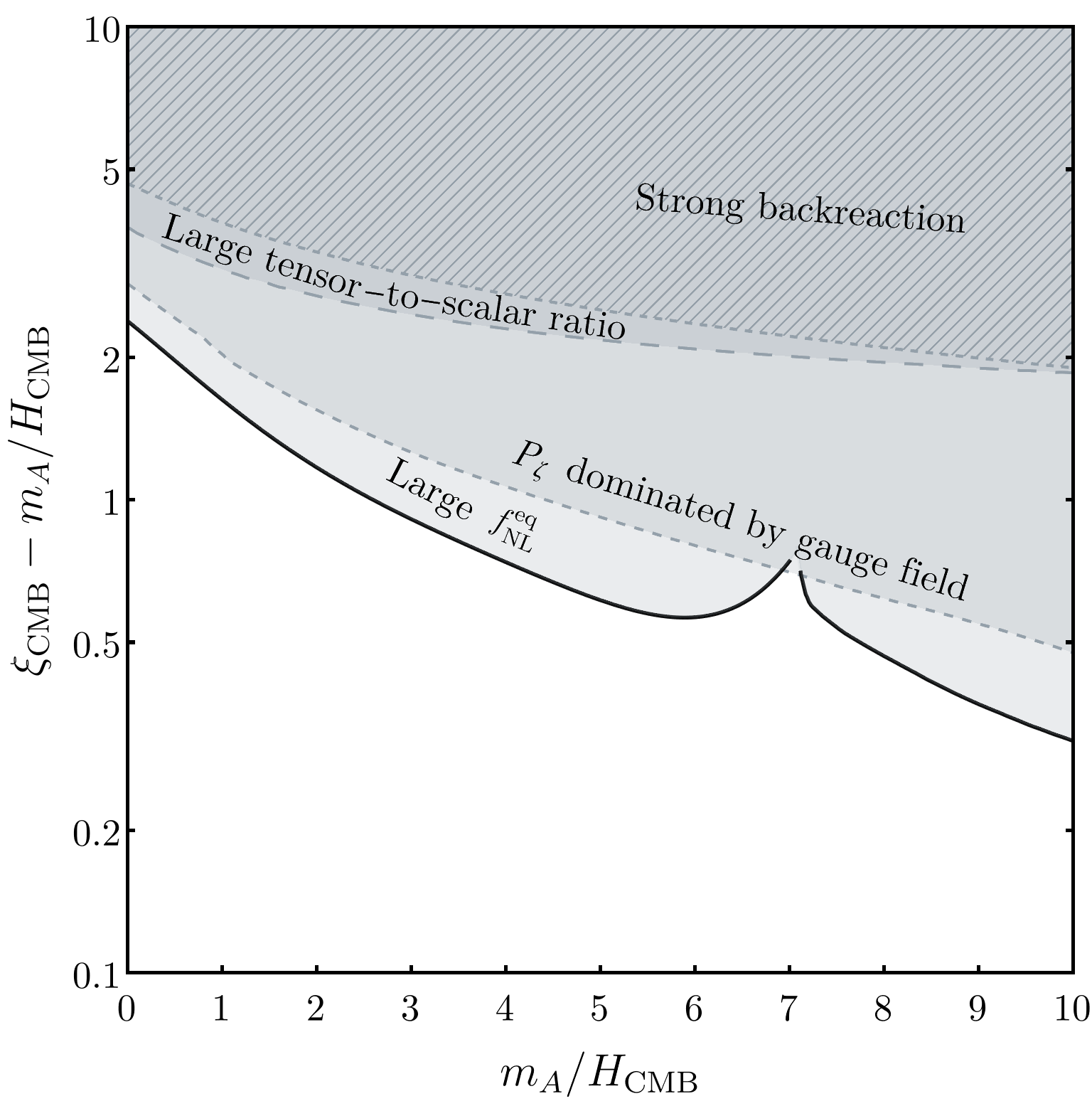}
    \caption{Shaded regions denote exclusion of the gauge boson's parameter space from various constraints. Tensor-to-scalar ratio bound is drawn for $H/M_{\rm Pl} = 10^{-5}$. For $f_{\rm NL}^{\rm eq} < -25\pm 47$ \cite{Planck:2019kim}, the left (right) part corresponds to the positive (negative) bound.} \label{fig:AllConstraints}
\end{figure}

Gauge field production during inflation may introduce significant non-Gaussianity in the curvature perturbations. It can be parametrized by a dimensionless quantity,
\begin{align}
    f_{\rm NL}^{\rm eq} &= \frac{10}{9 } \frac{k^6 }{ (2\pi)^{4}} \frac{\langle\zeta_{\mathbf{k_1}} \zeta_{\mathbf{k_2}} \zeta_{\mathbf{k_3}}\rangle'}{P_\zeta(k)^2 },
\end{align}
where the superscript `eq' represents the equilateral shape $k_1 = k_2 = k_3 \equiv k$ of the three-point correlation function. Current bound on equilateral non-Gaussianity gives $f_{\rm NL}^{\rm eq} = -25 \pm 47$ at $68\%$ CL \cite{Planck:2019kim}.


Another important constraint comes from the ratio of the tensor to scalar power, $r \equiv P_h / P_\zeta$  at the CMB scales, where the tensor power spectrum is chiral and can be decomposed as
\begin{align}
    P_h = \left[\frac{1}{\pi^2} \left( \frac{H}{M_{\rm Pl}} \right)^2 + P_h^{[A],+}\right] + \left[\frac{1}{\pi^2} \left( \frac{H}{M_{\rm Pl}} \right)^2 + P_h^{[A],-}\right] 
    &= P_h^+ + P_h^-, \label{PtensorTotal}
\end{align}
where $\pm$ corresponds to the two polarizations of the graviton. Here the vacuum contribution has been equally distributed to the two polarizations. In our case $P_h^+ \gg P_h^-$, implying a {parity violating} tensor power spectrum.  At the CMB scale, tensor-to-scalar ratio is constrained by current data $r_* \leq 0.056$ at $95\%$ CL \cite{Planck:2018jri}. 

In fig.~\ref{fig:AllConstraints}, we show the parameter space violating the above constraints, along with the region where the assumptions of negligible backreaction and negligible contribution of the gauge modes to the scalar power spectrum at CMB scales breaks down.

Evidently, the upper bound on scalar non-Gaussianity puts the strictest constraint on the parameter space of the model at this stage. It implies that $\xi$ is restricted not too far from $m_A/H$ at the CMB scale. However, as we will show in the next section, even for $\xi-m_A/H \sim \mc O(1)$ at the CMB scales, the power spectrum of tensor perturbations can evolve to generate observable GW spectrum at PTA and other interferometer scales.

\section{Evolution beyond the CMB scale}
Modes responsible for CMB scale observables can be assumed to experience constant $\xi$ and $H$ as long as slow-roll conditions prevail. However, modes contributing to the GW spectrum at smaller scales experience the time evolution of these parameters and are subject to strong backreaction from the inverse decay of the gauge field. These effects can be studied from a simultaneous solution of the coupled \cref{backr1,backr2}, ignoring the source term in \cref{backr2} as it is negligible compared to the source term in \cref{backr1}.
For convenience, we change variables from time $t$ to the e-folding number $N$, defined as $\mathrm{d}N = -H \mathrm{d}t$, where $N$ decreases as we approach the end of inflation. Eqs. \eqref{backr1} and \eqref{backr2} can then be expressed as
\begin{gather}
    \frac{\mathrm{d}^2\phi}{\mathrm{d}N^2} + \frac{\mathrm{d}\phi}{\mathrm{d}N} \left( 3+\frac{\mathrm{d}\ \log{H}}{\mathrm{d}N} \right) + \frac{1}{H^2} \frac{\mathrm{d}V}{\mathrm{d}\phi} = \frac{1}{H^2} \frac{1}{\Lambda} \langle \mathbf{E} \cdot \mathbf{B} \rangle, \label{phiEOM} \\
    \frac{1}{H^2} \approx \frac{1}{V}\ \left[3-\frac{1}{2}\left( \frac{\mathrm{d}\phi}{\mathrm{d}N} \right)^2\right]. \label{HEOM}
\end{gather}
Solving eqs.~\eqref{phiEOM} and \eqref{HEOM} numerically for a given potential $V$, we get $H(N)$ and $\phi(N)$, which can be used to calculate $\xi(N) \equiv \frac{1}{2\Lambda}\frac{d\phi}{dN} $.

\begin{figure}
    \centering
    \includegraphics[width=0.55\textwidth]{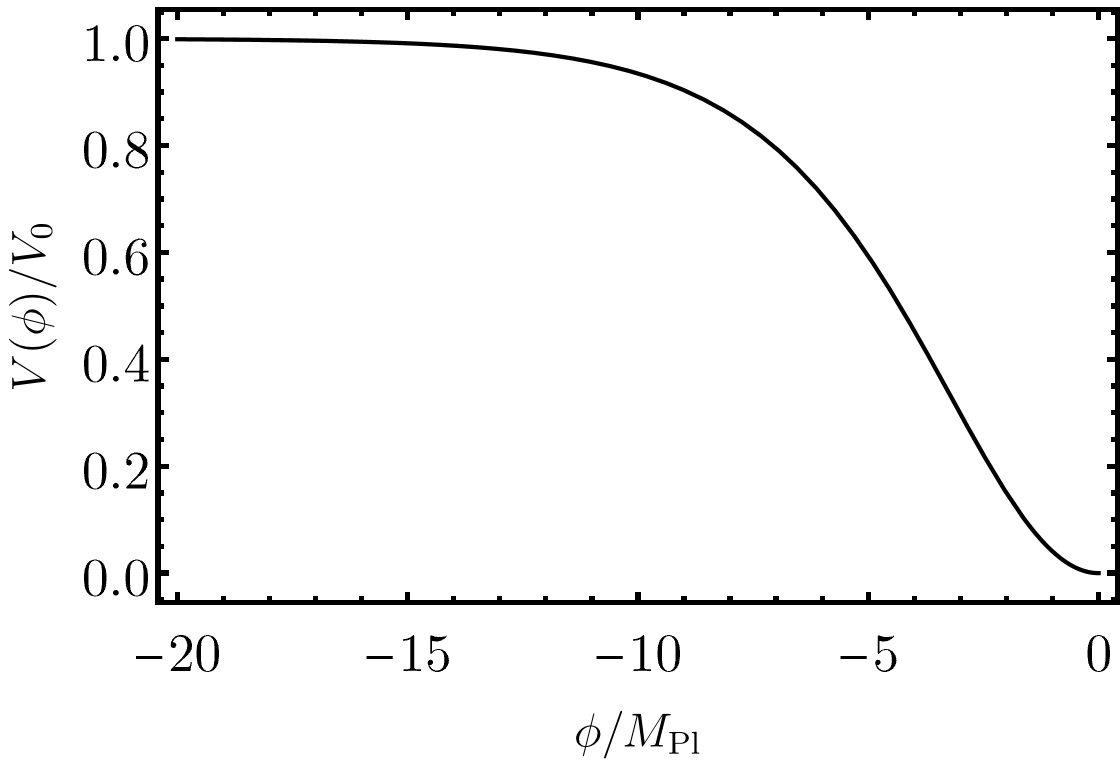}
    \caption{Inflaton potential for the T-model in \cref{Tmodel}, setting $\alpha_T = 4$.} \label{fig:Potential}
\end{figure}

\begin{figure}
    \centering
    \includegraphics[width=0.99\textwidth]{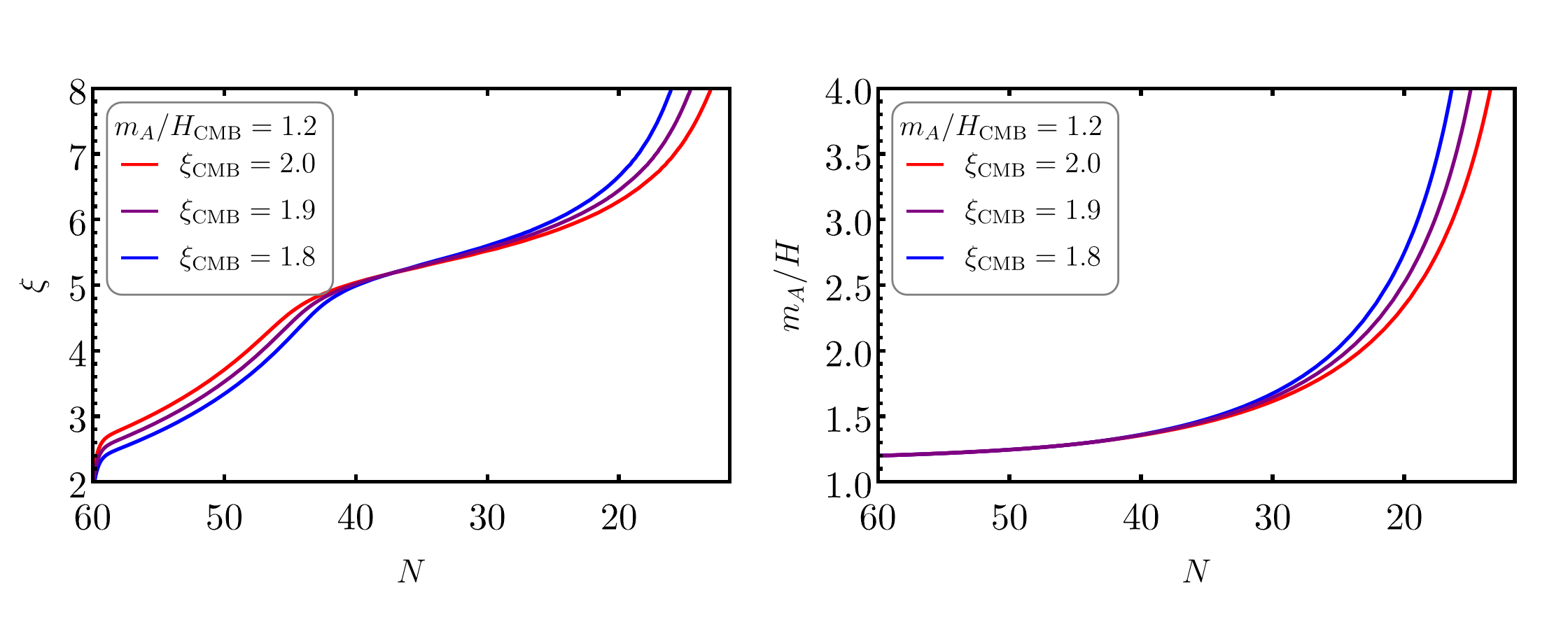} 
    \caption{Evolution of model parameters $\xi$ and $m_A/H$ for three benchmark points  in the context of the T-model inflaton potential. }
    \label{fig:evolwithN}
\end{figure}

We now specialize to the example of the T-model potential, which yields a consistent combination of the spectral index, $n_s$, and the tensor-to-scalar ratio, $r$ with respect to the combined Planck 2018 data analysis \cite{Planck:2018jri}. The inflaton potential in this model is given by
\begin{align}
    V(\phi) &= V_0 \tanh^{2}\left[ \frac{\phi}{\sqrt{6\alpha_T}M_{\rm Pl}} \right], \label{Tmodel}
\end{align}
where $V_0$ and $\alpha_T$ are free parameters, which can be constrained from CMB measurements of $n_s = 0.9649 \pm 0.0042$ (at $68\%$ CL) and $r < 0.056$ (at $95\%$ CL) \cite{Planck:2018jri}. 
We choose $\alpha_T = 4$ and $V_0 \approx 1.7 \times 10^{-9}$ which are consistent with all of the above constraints. The shape of the potential is shown in \cref{fig:Potential}. 
We illustrate the evolution of $\xi$ and $m_A/H$ as a function of $N$ in \cref{fig:evolwithN} for three benchmark points for this potential. 

At $N\sim 60$ when the CMB modes leave the horizon, standard slow-roll condition prevails and backreaction effects can be neglected. Furthermore, the selected benchmark points are in the allowed region in fig.~\ref{fig:AllConstraints}, so that large non-Gaussianities can be avoided. Beyond that, $\xi$ increases rapidly for about 20 e-folds after which backreaction effects slow down its rise. Closer to the end of inflation, slow-roll condition is re-established, and $\xi$ rises rapidly. In this regime, however, the Hubble rate also rises swiftly, compensating the parameter $m_A/H$ with respect to $\xi$, so that particle production remains under control.

\section{Gravitational wave signatures} \label{sec:7}
We now turn to the gravitational wave spectrum of the model potentially observed at NANOGrav and within the sensitivity of various upcoming interferometers.
The tensor perturbations sourced by the gauge modes leave the horizon during inflation and can source gravitational waves after re-entering the horizon at later stage. Taking the redshift into account, the amplitude of the GW signal observed today is given by
\begin{align}
    \Omega_{\rm GW}(f) &\equiv \frac{1}{24}\Omega_{R,0} P_h(f). \label{OmegaGW}
\end{align}
Here $\Omega_{R,0} \simeq 8.6 \times 10^{-5}$ denotes the radiation energy density today and $P_h(f)$ is the frequency dependent power spectrum of the tensor fluctuations at the time of horizon exit. Since $P_h^+ \gg P_h^-$, we have a parity violating GW spectrum. 

The power spectrum depends on the model parameters $\xi$ and $m_A/H$, whose time evolution was discussed in sec. 3. We can relate the time parameter $N$ to frequency of the GW signal $f$ as \cite{Domcke:2016bkh}
\begin{align}
    N = N_{\rm CMB} + \log{\frac{k_{\rm CMB}}{0.002\ \rm{Mpc}^{-1}}} - 19.57 - \log{\frac{f}{10^{-9}\ \rm{Hz}}}, \label{Nandf}
\end{align}
where the CMB pivot scale is $k_{\rm CMB} = 0.002\ \rm{Mpc}^{-1}$ and we take $N_{\rm CMB} = 60$.

Here we specifically focus on the parameter space that can potentially explain the observed excess at the NANOGrav 15-yr data. The effect of the gauge field creation on the tensor fluctuations is minimal for CMB scales and the power spectrum is dominated by the vacuum fluctuations. Current bound on scale-invariant stochastic gravitational wave at the CMB scales \cite{Namikawa:2019tax} implies a tensor-to-scalar ratio $r < 0.056$ \cite{Planck:2018jri}, which gives $H/M_{\rm Pl} \lesssim 2.6 \times 10^{-5}$, and $\Omega_{\rm GW} < 1.2 \times 10^{-16}$. 
Larger frequencies correspond to modes which left the horizon later than the CMB modes. By that time the rolling speed of the inflaton increases and the Hubble rate decreases, the combined effect of which implies a larger value of $\xi$. This dramatically enhances the power spectrum of the tensor perturbations sourced by the gauge field and it quickly supersedes the contribution from the vacuum fluctuations. Gravitational wave amplitude that eludes observation at the CMB scale now offers the possibility of detection at the interferometer scales. 

Although the blue-tilting of the tensor power spectrum at higher frequencies is a known feature of the axion inflation scenario, achieving a signal at the nano-Hz range probed by PTAs is significantly challenging. It requires the signal to start to rise at sufficiently smaller frequencies, yet such signals should not violate the upper bound at the LIGO-VIRGO-KAGRA (LVK) \cite{LIGOScientific:2022sts, KAGRA:2021kbb, Jiang:2022uxp}. Furthermore, such a wideband signal with a blue-tilted spectrum is likely to introduce significant contribution to $\Delta N_{\rm eff}$. These characteristics depend essentially on the inflaton potential one considers, as the rolling speed of the inflaton is determined by the potential. Intriguingly, we find that such peculiar features can be nicely accommodated in the T-model potential, but are difficult to achieve in the broader class of $\alpha$-attractor models \cite{Galante:2014ifa, Kallosh:2013maa, Kallosh:2013pby, Kallosh:2013yoa, Kallosh:2013hoa, Kallosh:2013lkr}. 

\begin{figure} \label{fig:PTArcade}
    \centering
    \includegraphics[width=0.76\textwidth]{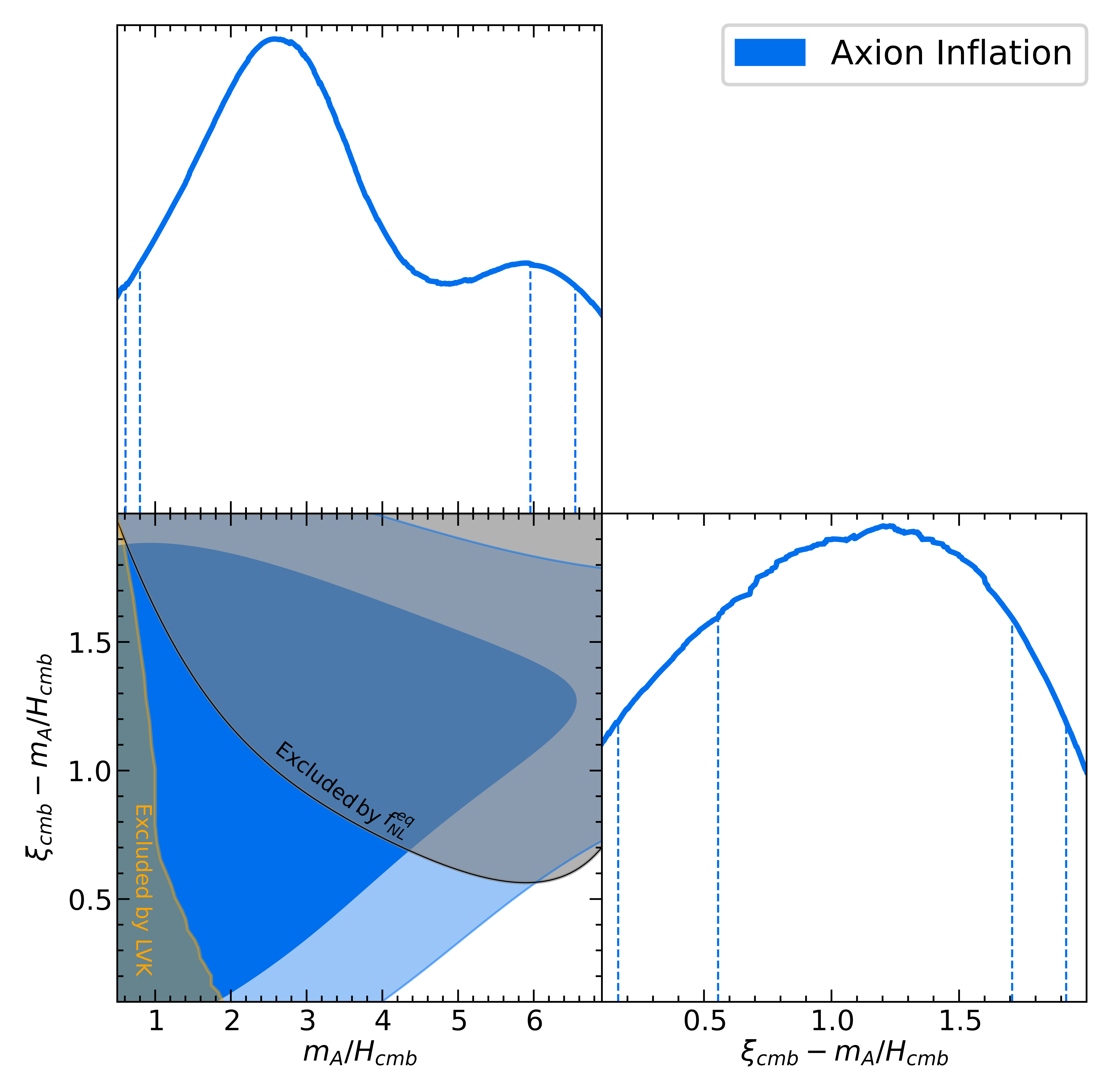}
    \caption{Posterior distributions for the axion inflation model parameters $m_A/H_{\rm CMB}$ and $\xi_{\rm CMB} - m_A/H_{\rm CMB}$. The diagonal subplots represent 1D marginalized distributions  with vertical lines showing $68\%$ and $95\%$ Bayesian credible intervals. In the lower left subplot, the darker and lighter blue regions show the $68\%$ and $95\%$ Bayesian credible regions in the 2D posterior distributions. The gray region is excluded by the upper bound on equilateral $f_{\rm NL}$. The orange region shows the parameter space excluded by the upper bound on GW amplitude at LVK. 
    }
\end{figure}

We show the posterior distributions of the model parameters $m_A/H_{\rm CMB}$ and $\xi_{\rm CMB} - m_A/H_{\rm CMB}$ with dashed lines, and $68\%$ (darker blue) and $95\%$ (lighter blue) Bayesian credible regions for the NANOGrav 15yr results using \texttt{PTArcade} \cite{Mitridate:2023oar} in fig.~\ref{fig:PTArcade}. The 2D posterior regions shrink for larger $m_A/H_{\rm CMB}$ as backreaction from gauge fields becomes so strong for larger $\xi_{\rm CMB}$ that the amplitude of the GW signal at NANOGrav frequencies is diminished. In the same plot we also show the constraints from $f_{\rm NL}$ in the equilateral limit (gray region) and excess GW amplitude at LVK scales (orange region). Interestingly, these constraints completely rule out the parameter space for $m_A/H \lesssim 0.6$. 

\begin{figure} 
    \centering
    \includegraphics[width=0.99\textwidth]{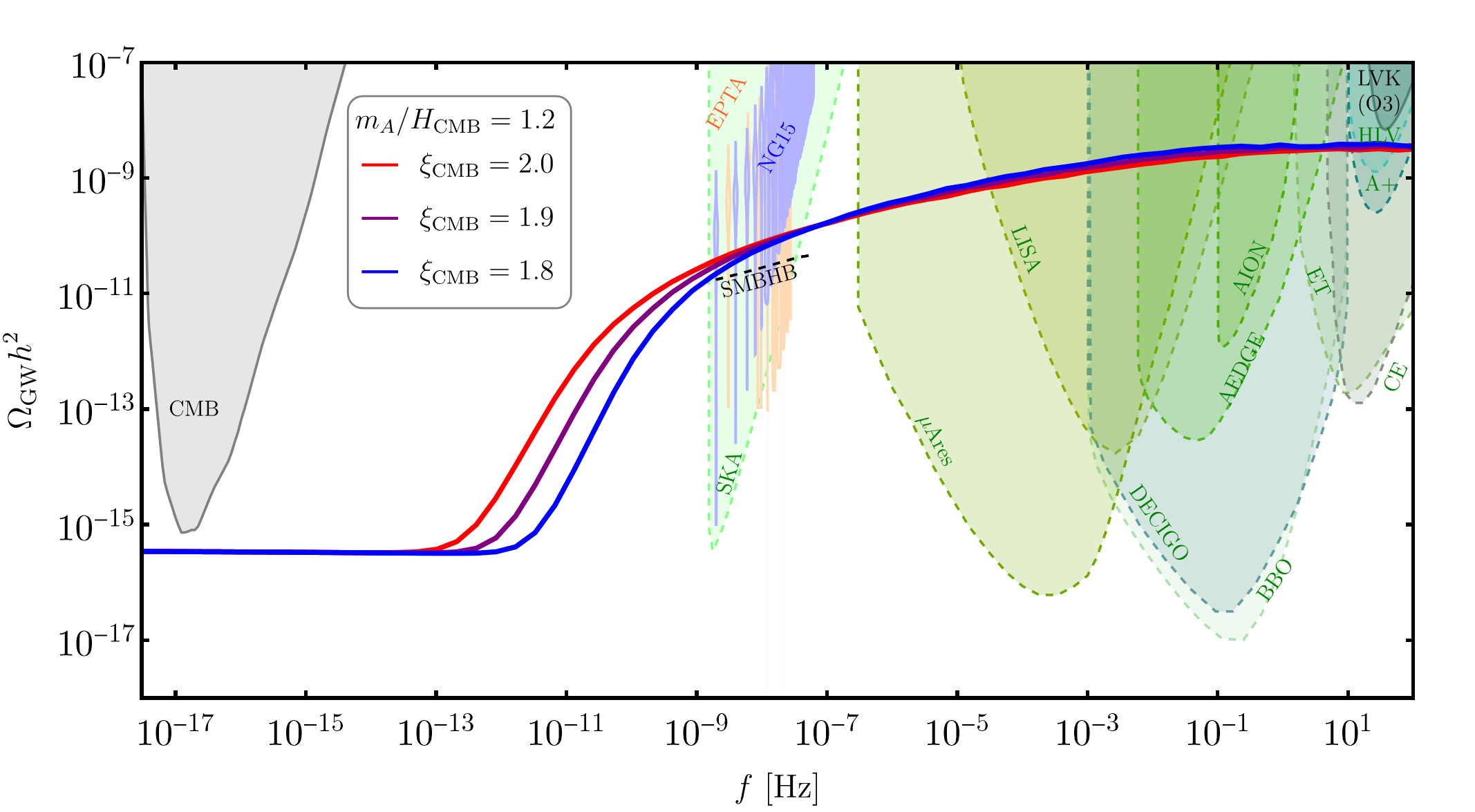}
    \caption{Gravitational wave spectrum for gauge boson production in the context of the T-model potential. 
    }\label{fig:GWdetail2}
\end{figure}

We show the GW spectrum for three benchmark points from the viable parameter space in fig.~\ref{fig:GWdetail2}. In each case we keep the parameter $m_A/H$ same at the CMB scales but let the parameter $\xi$ vary. For comparison we show the upper limits from CMB \cite{Namikawa:2019tax} and LVK \cite{LIGOScientific:2022sts, KAGRA:2021kbb, Jiang:2022uxp}, and the projected sensitivites of future detectors SKA \cite{Janssen:2014dka}, $\mu$-Ares \cite{Sesana:2019vho}, LISA \cite{LISA:2017pwj}, DECIGO \cite{Kudoh:2005as, Kawamura:2020pcg}, BBO \cite{Harry:2006fi}, AEDGE \cite{AEDGE:2019nxb}, AION \cite{Badurina:2019hst}, CE \cite{LIGOScientific:2016wof}, ET \cite{Hild:2008ng} and future upgrades of LVK. The recent results from NANOGrav and EPTA are shown with blue and orange violins, and the expected astrophysical background from inspiraling supermassive black hole binaries (SMBHBs) \cite{NANOGrav:2023hvm, NANOGrav:2023hfp} is shown in the PTA frequencies with a black dashed line. \textcolor{black}{The GW spectrum from SMBHB best fit parameters fall short both in amplitude and slope in fully explaining the power law fit of the observed GW background in the NANOGrav data, which could be improved if a GW signal with larger amplitude and steeper slope were present \cite{NANOGrav:2023hvm}. All the three benchmark signals originating from our model satisfy this criteria. This leads to an improved fit to the spectral characteristics of the NANOGrav signal.} For all three benchmark points, the signal remains flat at smaller frequencies as the contribution from gauge boson production remains subdominant compared to the vacuum fluctuations. The signals begin to rise at $f \gtrsim 10^{-13}$ Hz, but the growth is slowed down near the PTA frequencies. For higher frequencies, backreaction effects saturate the signals so that they remain below the upper bound at LVK frequencies. These signals can be detected in various terrestrial and space-based interferometers from nano-Hz to kilo-Hz range.

While wideband stochastic GW signals can appear from some other sources, for example, topological defects like cosmic strings, an important distinction of the signals produced in the present case is their {parity-violating} nature. Such parity-violating nature is expected to be detected in the ET-CE \cite{Badger:2021enh} and the LISA-Taiji network \cite{Orlando:2020oko}.

We now calculate the contribution of the GW spectrum to the radiation energy budget of the Universe in terms of the effective number of relativistic species $N_{\rm eff}$. The extra contribution is given by
\begin{align}
    \Delta N_{\rm eff} \simeq 1.8 \times 10^{5} \int_{f_{\rm min}}^{f_{\rm max}} df \frac{\Omega_{\rm GW}(f) h^2}{f},
\end{align}
where $f_{\rm min}$ and $f_{\rm max}$ depend on the era of interest and the maximum temperature reached. One needs to make sure that the GW spectrum does not spoil the successful prediction of BBN, corresponding to the frequency of a mode crossing the horizon during the time of BBN when the temperature
of the radiation bath was $T \sim  \mc O({\rm MeV})$. We take $f_{\rm max} = 10^4$ Hz for a sufficiently large reheating temperature. The current  upper bound on $\Delta N_{\rm eff}$ from BBN and CMB probes is $\Delta N_{\rm eff} \simeq 0.4$ \cite{Planck:2018vyg}. We find that all the benchmark points satisfy this bound, with $\Delta N_{\rm eff} \sim 0.013$ in all three cases. In fact, we can do a naive estimate to show that any signal that does not violate the LVK bound would also not violate the $\Delta N_{\rm eff}$ bound in our model. Since the spectrum does not have any peak, suppose we take it to be flat at $\Omega_{\rm GW}h^2 \sim 10^{-8}$ for all relevant frequencies. Even then, we find $\Delta N_{\rm eff} \lesssim 0.1$, well below the upper bound from BBN and CMB probes. 

\begin{figure}
    \centering
    \includegraphics[width=0.69\textwidth]{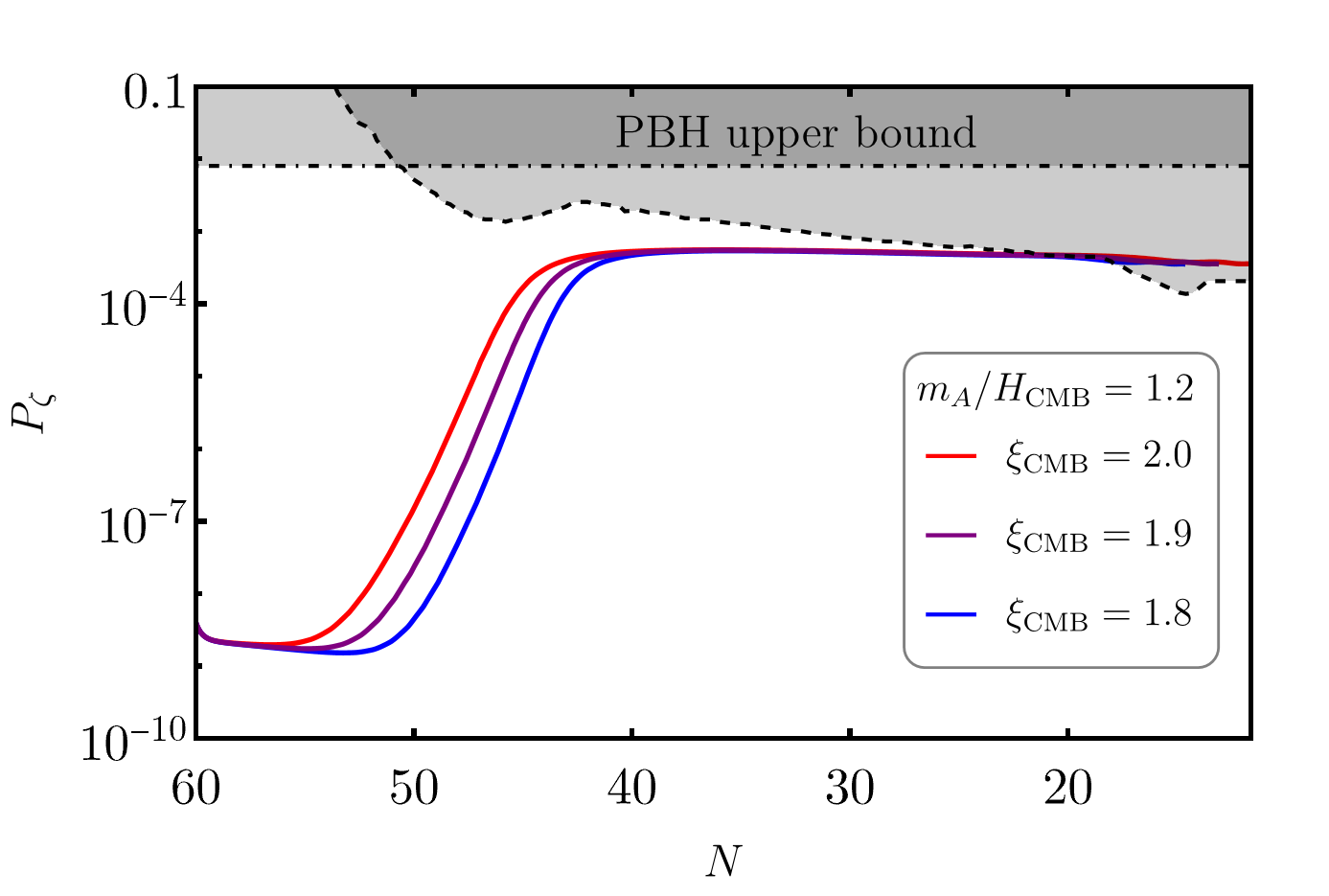}
    \caption{Scalar power spectrum for the T-model potential. Gray regions represent upper bounds from the overproduction of primordial black holes.}
    \label{fig:PScalarBR}
\end{figure}

Both the tensor and scalar perturbations are enhanced by gauge fields. Large scalar perturbations can lead to the creation of primordial black holes (PBH). 
The mass of a PBH is related to the e-folding number $N$ when a fluctuation responsible for the creation of the PBH leaves the horizon. An upper bound on the scalar power spectrum as a function of $N$ has been derived in ref.~\cite{Linde:2012bt} using the estimates of refs.~\cite{Josan:2009qn, Carr:2020gox}, as shown in \cref{fig:PScalarBR} with a dashed curve. In the same plot we show the scalar power spectrum for the three benchmark points. In all cases the power spectrum suffers from strong backreaction at smaller scales. Nevertheless, we see some tension of the spectrum with the PBH bound, which is within the theoretical uncertainties. Furthermore, recent lattice simulations \cite{Caravano:2022epk} indicate that the combined effect of non-Gaussian perturbations become more Gaussian overall at smaller scales, which weakens the bound (shown with a dot-dashed line in \cref{fig:PScalarBR}) and allows the model to avoid overproduction of PBHs. The problem can be further alleviated by considering $N$ copies of the gauge field, which reduces the scalar power spectrum by a factor of $N$ \cite{Anber:2009ua}, but does not significantly affect the tensor power spectrum at smaller scales. Large scalar perturbations can also generate second order tensor perturbations \cite{Baumann:2007zm}, which will be discussed in a future work.


\section{Conclusion and outlook} \label{sec:8}
We have pointed out that an explanation of the observed excess in the NANOGrav 15-yr dataset is possible from the gravitational waves generated in axion inflation from axion coupling to (massive) gauge bosons. Such a coupling, natural for a  shift-symmetric inflaton, leads to copious particle production during inflation, leaving an indelible imprint on the primordial scalar and tensor perturbations. This leads to a unique {parity-violating} contribution to the GW spectrum, that remains flat at CMB scales, but blue-tilts at smaller scales and can become audible to pulsar timing arrays. The growth of the spectrum at higher frequencies, potentially dangerous for LVK scales, is controlled by the backreaction of the gauge quanta on the inflationary dynamics, the details of which depend somewhat on the inflaton potential. We have specifically discussed the case of the T-model potential, a model currently favored by data, and have identified the parameter space that can potentially produce a GW spectrum statistically consistent with the NANOGrav result. 
More specifically, we have shown three benchmark points for which the amplitude and spectral slope of the GW signals provide better fit to the NANOGrav result than the standard astrophysical background from inspiraling SMBHBs, suggesting an interpretation of the NANOGrav excess from GW waves generated in axion inflation is quite likely. The GW spectrum can also be probed in various detectors from nano-Hz to kilo-Hz frequencies, while its {parity-violating} nature would clearly distinguish it from other wideband signals.

\acknowledgments
We have benefitted from useful discussions with Pasquale Di Bari, Anish Ghoshal, Caner Ünal and Wei Xue. X.N. is grateful for the technical support from Yuxin Zhao. X.N. is partially supported by the U.S. Department of Energy under grant DE-SC0022148 at the University of Florida. M.H.R. acknowledges  support from the STFC Consolidated Grant ST/T000775/1, and from the European Union's Horizon 2020 Research and Innovation
Programme under Marie Sklodowska-Curie grant agreement HIDDeN European
ITN project (H2020-MSCA-ITN-2019//860881-HIDDeN).

\bibliography{references}

\providecommand{\href}[2]{#2}\begingroup\raggedright\begin{thebibliography}{100}

\bibitem{LIGOScientific:2016aoc}
{\scshape LIGO Scientific, Virgo} collaboration, \emph{{Observation of
  Gravitational Waves from a Binary Black Hole Merger}},
  \href{https://doi.org/10.1103/PhysRevLett.116.061102}{\emph{Phys. Rev. Lett.}
  {\bfseries 116} (2016) 061102}
  [\href{https://arxiv.org/abs/1602.03837}{{\ttfamily 1602.03837}}].

\bibitem{EPTA:2023sfo}
{\scshape EPTA} collaboration, \emph{{The second data release from the European
  Pulsar Timing Array I. The dataset and timing analysis}},
  \href{https://arxiv.org/abs/2306.16224}{{\ttfamily 2306.16224}}.

\bibitem{EPTA:2023akd}
{\scshape EPTA} collaboration, \emph{{The second data release from the European
  Pulsar Timing Array II. Customised pulsar noise models for spatially
  correlated gravitational waves}},
  \href{https://arxiv.org/abs/2306.16225}{{\ttfamily 2306.16225}}.

\bibitem{EPTA:2023fyk}
{\scshape EPTA} collaboration, \emph{{The second data release from the European
  Pulsar Timing Array III. Search for gravitational wave signals}},
  \href{https://arxiv.org/abs/2306.16214}{{\ttfamily 2306.16214}}.

\bibitem{EPTA:2023gyr}
{\scshape EPTA} collaboration, \emph{{The second data release from the European
  Pulsar Timing Array IV. Search for continuous gravitational wave signals}},
  \href{https://arxiv.org/abs/2306.16226}{{\ttfamily 2306.16226}}.

\bibitem{EPTA:2023xxk}
{\scshape EPTA} collaboration, \emph{{The second data release from the European
  Pulsar Timing Array: V. Implications for massive black holes, dark matter and
  the early Universe}},  \href{https://arxiv.org/abs/2306.16227}{{\ttfamily
  2306.16227}}.

\bibitem{EPTA:2023xiy}
{\scshape EPTA} collaboration, \emph{{The second data release from the European
  Pulsar Timing Array: VI. Challenging the ultralight dark matter paradigm}},
  \href{https://arxiv.org/abs/2306.16228}{{\ttfamily 2306.16228}}.

\bibitem{Reardon:2023gzh}
D.J.~Reardon et~al., \emph{{Search for an Isotropic Gravitational-wave
  Background with the Parkes Pulsar Timing Array}},
  \href{https://doi.org/10.3847/2041-8213/acdd02}{\emph{Astrophys. J. Lett.}
  {\bfseries 951} (2023) L6}
  [\href{https://arxiv.org/abs/2306.16215}{{\ttfamily 2306.16215}}].

\bibitem{Reardon:2023zen}
D.J.~Reardon et~al., \emph{{The Gravitational-wave Background Null Hypothesis:
  Characterizing Noise in Millisecond Pulsar Arrival Times with the Parkes
  Pulsar Timing Array}},
  \href{https://doi.org/10.3847/2041-8213/acdd03}{\emph{Astrophys. J. Lett.}
  {\bfseries 951} (2023) L7}
  [\href{https://arxiv.org/abs/2306.16229}{{\ttfamily 2306.16229}}].

\bibitem{Zic:2023gta}
A.~Zic et~al., \emph{{The Parkes Pulsar Timing Array Third Data Release}},
  \href{https://arxiv.org/abs/2306.16230}{{\ttfamily 2306.16230}}.

\bibitem{Antoniadis:2022pcn}
J.~Antoniadis et~al., \emph{{The International Pulsar Timing Array second data
  release: Search for an isotropic gravitational wave background}},
  \href{https://doi.org/10.1093/mnras/stab3418}{\emph{Mon. Not. Roy. Astron.
  Soc.} {\bfseries 510} (2022) 4873}
  [\href{https://arxiv.org/abs/2201.03980}{{\ttfamily 2201.03980}}].

\bibitem{Xu:2023wog}
H.~Xu et~al., \emph{{Searching for the Nano-Hertz Stochastic Gravitational Wave
  Background with the Chinese Pulsar Timing Array Data Release I}},
  \href{https://doi.org/10.1088/1674-4527/acdfa5}{\emph{Res. Astron.
  Astrophys.} {\bfseries 23} (2023) 075024}
  [\href{https://arxiv.org/abs/2306.16216}{{\ttfamily 2306.16216}}].

\bibitem{NANOGrav:2023gor}
{\scshape NANOGrav} collaboration, \emph{{The NANOGrav 15 yr Data Set: Evidence
  for a Gravitational-wave Background}},
  \href{https://doi.org/10.3847/2041-8213/acdac6}{\emph{Astrophys. J. Lett.}
  {\bfseries 951} (2023) L8}
  [\href{https://arxiv.org/abs/2306.16213}{{\ttfamily 2306.16213}}].

\bibitem{NANOGrav:2023hvm}
{\scshape NANOGrav} collaboration, \emph{{The NANOGrav 15 yr Data Set: Search
  for Signals from New Physics}},
  \href{https://doi.org/10.3847/2041-8213/acdc91}{\emph{Astrophys. J. Lett.}
  {\bfseries 951} (2023) L11}
  [\href{https://arxiv.org/abs/2306.16219}{{\ttfamily 2306.16219}}].

\bibitem{NANOGrav:2023hfp}
{\scshape NANOGrav} collaboration, \emph{{The NANOGrav 15-year Data Set:
  Constraints on Supermassive Black Hole Binaries from the Gravitational Wave
  Background}},  \href{https://arxiv.org/abs/2306.16220}{{\ttfamily
  2306.16220}}.

\bibitem{NANOGrav:2023icp}
{\scshape NANOGrav} collaboration, \emph{{The NANOGrav 15-year
  Gravitational-Wave Background Analysis Pipeline}},
  \href{https://arxiv.org/abs/2306.16223}{{\ttfamily 2306.16223}}.

\bibitem{NANOGrav:2023ctt}
{\scshape NANOGrav} collaboration, \emph{{The NANOGrav 15 yr Data Set: Detector
  Characterization and Noise Budget}},
  \href{https://doi.org/10.3847/2041-8213/acda88}{\emph{Astrophys. J. Lett.}
  {\bfseries 951} (2023) L10}
  [\href{https://arxiv.org/abs/2306.16218}{{\ttfamily 2306.16218}}].

\bibitem{NANOGrav:2023hde}
{\scshape NANOGrav} collaboration, \emph{{The NANOGrav 15 yr Data Set:
  Observations and Timing of 68 Millisecond Pulsars}},
  \href{https://doi.org/10.3847/2041-8213/acda9a}{\emph{Astrophys. J. Lett.}
  {\bfseries 951} (2023) L9}
  [\href{https://arxiv.org/abs/2306.16217}{{\ttfamily 2306.16217}}].

\bibitem{NANOGrav:2023tcn}
{\scshape NANOGrav} collaboration, \emph{{The NANOGrav 15-year Data Set: Search
  for Anisotropy in the Gravitational-Wave Background}},
  \href{https://arxiv.org/abs/2306.16221}{{\ttfamily 2306.16221}}.

\bibitem{Hellings:1983fr}
R.w.~Hellings and G.s.~Downs, \emph{{UPPER LIMITS ON THE ISOTROPIC
  GRAVITATIONAL RADIATION BACKGROUND FROM PULSAR TIMING ANALYSIS}},
  \href{https://doi.org/10.1086/183954}{\emph{Astrophys. J. Lett.} {\bfseries
  265} (1983) L39}.

\bibitem{Guth:1980zm}
A.H.~Guth, \emph{{The Inflationary Universe: A Possible Solution to the Horizon
  and Flatness Problems}},
  \href{https://doi.org/10.1103/PhysRevD.23.347}{\emph{Phys. Rev. D} {\bfseries
  23} (1981) 347}.

\bibitem{Lyth:1998xn}
D.H.~Lyth and A.~Riotto, \emph{{Particle physics models of inflation and the
  cosmological density perturbation}},
  \href{https://doi.org/10.1016/S0370-1573(98)00128-8}{\emph{Phys. Rept.}
  {\bfseries 314} (1999) 1}
  [\href{https://arxiv.org/abs/hep-ph/9807278}{{\ttfamily hep-ph/9807278}}].

\bibitem{Kinney:2003xf}
W.H.~Kinney, \emph{{Cosmology, inflation, and the physics of nothing}},
  \href{https://doi.org/10.1007/978-94-010-0076-5_5}{\emph{NATO Sci. Ser. II}
  {\bfseries 123} (2003) 189}
  [\href{https://arxiv.org/abs/astro-ph/0301448}{{\ttfamily
  astro-ph/0301448}}].

\bibitem{Baumann:2009ds}
D.~Baumann, \emph{{Inflation}},  in \emph{{Theoretical Advanced Study Institute
  in Elementary Particle Physics}: {Physics of the Large and the Small}},
  pp.~523--686, 2011, \href{https://doi.org/10.1142/9789814327183_0010}{DOI}
  [\href{https://arxiv.org/abs/0907.5424}{{\ttfamily 0907.5424}}].

\bibitem{Domenech:2021ztg}
G.~Dom\`enech, \emph{{Scalar Induced Gravitational Waves Review}},
  \href{https://doi.org/10.3390/universe7110398}{\emph{Universe} {\bfseries 7}
  (2021) 398} [\href{https://arxiv.org/abs/2109.01398}{{\ttfamily
  2109.01398}}].

\bibitem{Yuan:2021qgz}
C.~Yuan and Q.-G.~Huang, \emph{{A topic review on probing primordial black hole
  dark matter with scalar induced gravitational waves}},
  \href{https://arxiv.org/abs/2103.04739}{{\ttfamily 2103.04739}}.

\bibitem{Athron:2023xlk}
P.~Athron, C.~Bal\'azs, A.~Fowlie, L.~Morris and L.~Wu, \emph{{Cosmological
  phase transitions: from perturbative particle physics to gravitational
  waves}},  \href{https://arxiv.org/abs/2305.02357}{{\ttfamily 2305.02357}}.

\bibitem{Kibble:1976sj}
T.W.B.~Kibble, \emph{{Topology of Cosmic Domains and Strings}},
  \href{https://doi.org/10.1088/0305-4470/9/8/029}{\emph{J. Phys. A} {\bfseries
  9} (1976) 1387}.

\bibitem{Vilenkin:1981bx}
A.~Vilenkin, \emph{{Gravitational radiation from cosmic strings}},
  \href{https://doi.org/10.1016/0370-2693(81)91144-8}{\emph{Phys. Lett. B}
  {\bfseries 107} (1981) 47}.

\bibitem{Vilenkin:2000jqa}
A.~Vilenkin and E.P.S.~Shellard, \emph{{Cosmic Strings and Other Topological
  Defects}}, Cambridge University Press (7, 2000).

\bibitem{Buchmueller:2023nll}
O.~Buchmueller, J.~Ellis and U.~Schneider, \emph{{Large-Scale Atom
  Interferometry for Fundamental Physics}},
  \href{https://arxiv.org/abs/2306.17726}{{\ttfamily 2306.17726}}.

\bibitem{Broadhurst:2023tus}
T.~Broadhurst, C.~Chen, T.~Liu and K.-F.~Zheng, \emph{{Binary Supermassive
  Black Holes Orbiting Dark Matter Solitons: From the Dual AGN in UGC4211 to
  NanoHertz Gravitational Waves}},
  \href{https://arxiv.org/abs/2306.17821}{{\ttfamily 2306.17821}}.

\bibitem{Yang:2023aak}
J.~Yang, N.~Xie and F.P.~Huang, \emph{{Nano-Hertz stochastic gravitational wave
  background as hints of ultralight axion particles}},
  \href{https://arxiv.org/abs/2306.17113}{{\ttfamily 2306.17113}}.

\bibitem{Yang:2023qlf}
A.~Yang, J.~Ma, S.~Jiang and F.P.~Huang, \emph{{Implication of nano-Hertz
  stochastic gravitational wave on dynamical dark matter through a first-order
  phase transition}},  \href{https://arxiv.org/abs/2306.17827}{{\ttfamily
  2306.17827}}.

\bibitem{Eichhorn:2023gat}
A.~Eichhorn, R.R.~Lino~dos Santos and J.a.L.~Miqueleto, \emph{{From quantum
  gravity to gravitational waves through cosmic strings}},
  \href{https://arxiv.org/abs/2306.17718}{{\ttfamily 2306.17718}}.

\bibitem{Huang:2023chx}
H.-L.~Huang, Y.~Cai, J.-Q.~Jiang, J.~Zhang and Y.-S.~Piao, \emph{{Supermassive
  primordial black holes in multiverse: for nano-Hertz gravitational wave and
  high-redshift JWST galaxies}},
  \href{https://arxiv.org/abs/2306.17577}{{\ttfamily 2306.17577}}.

\bibitem{Wang:2023ost}
S.~Wang, Z.-C.~Zhao, J.-P.~Li and Q.-H.~Zhu, \emph{{Exploring the Implications
  of 2023 Pulsar Timing Array Datasets for Scalar-Induced Gravitational Waves
  and Primordial Black Holes}},
  \href{https://arxiv.org/abs/2307.00572}{{\ttfamily 2307.00572}}.

\bibitem{Gouttenoire:2023ftk}
Y.~Gouttenoire and E.~Vitagliano, \emph{{Domain wall interpretation of the PTA
  signal confronting black hole overproduction}},
  \href{https://arxiv.org/abs/2306.17841}{{\ttfamily 2306.17841}}.

\bibitem{DiBari:2023upq}
P.~Di~Bari and M.H.~Rahat, \emph{{The split majoron model confronts the
  NANOGrav signal}},  \href{https://arxiv.org/abs/2307.03184}{{\ttfamily
  2307.03184}}.

\bibitem{Cai:2023dls}
Y.-F.~Cai, X.-C.~He, X.~Ma, S.-F.~Yan and G.-W.~Yuan, \emph{{Limits on
  scalar-induced gravitational waves from the stochastic background by pulsar
  timing array observations}},
  \href{https://arxiv.org/abs/2306.17822}{{\ttfamily 2306.17822}}.

\bibitem{Inomata:2023zup}
K.~Inomata, K.~Kohri and T.~Terada, \emph{{The Detected Stochastic
  Gravitational Waves and Sub-Solar Primordial Black Holes}},
  \href{https://arxiv.org/abs/2306.17834}{{\ttfamily 2306.17834}}.

\bibitem{Lazarides:2023ksx}
G.~Lazarides, R.~Maji and Q.~Shafi, \emph{{Superheavy quasi-stable strings and
  walls bounded by strings in the light of NANOGrav 15 year data}},
  \href{https://arxiv.org/abs/2306.17788}{{\ttfamily 2306.17788}}.

\bibitem{Depta:2023qst}
P.F.~Depta, K.~Schmidt-Hoberg and C.~Tasillo, \emph{{Do pulsar timing arrays
  observe merging primordial black holes?}},
  \href{https://arxiv.org/abs/2306.17836}{{\ttfamily 2306.17836}}.

\bibitem{Blasi:2023sej}
S.~Blasi, A.~Mariotti, A.~Rase and A.~Sevrin, \emph{{Axionic domain walls at
  Pulsar Timing Arrays: QCD bias and particle friction}},
  \href{https://arxiv.org/abs/2306.17830}{{\ttfamily 2306.17830}}.

\bibitem{Franciolini:2023wjm}
G.~Franciolini, D.~Racco and F.~Rompineve, \emph{{Footprints of the QCD
  Crossover on Cosmological Gravitational Waves at Pulsar Timing Arrays}},
  \href{https://arxiv.org/abs/2306.17136}{{\ttfamily 2306.17136}}.

\bibitem{Shen:2023pan}
Z.-Q.~Shen, G.-W.~Yuan, Y.-Y.~Wang and Y.-Z.~Wang, \emph{{Dark Matter Spike
  surrounding Supermassive Black Holes Binary and the nanohertz Stochastic
  Gravitational Wave Background}},
  \href{https://arxiv.org/abs/2306.17143}{{\ttfamily 2306.17143}}.

\bibitem{Zu:2023olm}
L.~Zu, C.~Zhang, Y.-Y.~Li, Y.-C.~Gu, Y.-L.S.~Tsai and Y.-Z.~Fan, \emph{{Mirror
  QCD phase transition as the origin of the nanohertz Stochastic
  Gravitational-Wave Background detected by the Pulsar Timing Arrays}},
  \href{https://arxiv.org/abs/2306.16769}{{\ttfamily 2306.16769}}.

\bibitem{Lambiase:2023pxd}
G.~Lambiase, L.~Mastrototaro and L.~Visinelli, \emph{{Astrophysical neutrino
  oscillations after pulsar timing array analyses}},
  \href{https://arxiv.org/abs/2306.16977}{{\ttfamily 2306.16977}}.

\bibitem{Han:2023olf}
C.~Han, K.-P.~Xie, J.M.~Yang and M.~Zhang, \emph{{Self-interacting dark matter
  implied by nano-Hertz gravitational waves}},
  \href{https://arxiv.org/abs/2306.16966}{{\ttfamily 2306.16966}}.

\bibitem{Guo:2023hyp}
S.-Y.~Guo, M.~Khlopov, X.~Liu, L.~Wu, Y.~Wu and B.~Zhu, \emph{{Footprints of
  Axion-Like Particle in Pulsar Timing Array Data and JWST Observations}},
  \href{https://arxiv.org/abs/2306.17022}{{\ttfamily 2306.17022}}.

\bibitem{Wang:2023len}
Z.~Wang, L.~Lei, H.~Jiao, L.~Feng and Y.-Z.~Fan, \emph{{The nanohertz
  stochastic gravitational-wave background from cosmic string Loops and the
  abundant high redshift massive galaxies}},
  \href{https://arxiv.org/abs/2306.17150}{{\ttfamily 2306.17150}}.

\bibitem{Ellis:2023tsl}
J.~Ellis, M.~Lewicki, C.~Lin and V.~Vaskonen, \emph{{Cosmic Superstrings
  Revisited in Light of NANOGrav 15-Year Data}},
  \href{https://arxiv.org/abs/2306.17147}{{\ttfamily 2306.17147}}.

\bibitem{Vagnozzi:2023lwo}
S.~Vagnozzi, \emph{{Inflationary interpretation of the stochastic gravitational
  wave background signal detected by pulsar timing array experiments}},
  \href{https://arxiv.org/abs/2306.16912}{{\ttfamily 2306.16912}}.

\bibitem{Fujikura:2023lkn}
K.~Fujikura, S.~Girmohanta, Y.~Nakai and M.~Suzuki, \emph{{NANOGrav Signal from
  a Dark Conformal Phase Transition}},
  \href{https://arxiv.org/abs/2306.17086}{{\ttfamily 2306.17086}}.

\bibitem{Kitajima:2023cek}
N.~Kitajima, J.~Lee, K.~Murai, F.~Takahashi and W.~Yin, \emph{{Nanohertz
  Gravitational Waves from Axion Domain Walls Coupled to QCD}},
  \href{https://arxiv.org/abs/2306.17146}{{\ttfamily 2306.17146}}.

\bibitem{Li:2023yaj}
Y.~Li, C.~Zhang, Z.~Wang, M.~Cui, Y.-L.S.~Tsai, Q.~Yuan et~al.,
  \emph{{Primordial magnetic field as a common solution of nanohertz
  gravitational waves and Hubble tension}},
  \href{https://arxiv.org/abs/2306.17124}{{\ttfamily 2306.17124}}.

\bibitem{Franciolini:2023pbf}
G.~Franciolini, A.~Iovino, Junior., V.~Vaskonen and H.~Veermae, \emph{{The
  recent gravitational wave observation by pulsar timing arrays and primordial
  black holes: the importance of non-gaussianities}},
  \href{https://arxiv.org/abs/2306.17149}{{\ttfamily 2306.17149}}.

\bibitem{Megias:2023kiy}
E.~Megias, G.~Nardini and M.~Quiros, \emph{{Pulsar Timing Array Stochastic
  Background from light Kaluza-Klein resonances}},
  \href{https://arxiv.org/abs/2306.17071}{{\ttfamily 2306.17071}}.

\bibitem{Ellis:2023dgf}
J.~Ellis, M.~Fairbairn, G.~H\"utsi, J.~Raidal, J.~Urrutia, V.~Vaskonen et~al.,
  \emph{{Gravitational Waves from SMBH Binaries in Light of the NANOGrav
  15-Year Data}},  \href{https://arxiv.org/abs/2306.17021}{{\ttfamily
  2306.17021}}.

\bibitem{Bai:2023cqj}
Y.~Bai, T.-K.~Chen and M.~Korwar, \emph{{QCD-Collapsed Domain Walls: QCD Phase
  Transition and Gravitational Wave Spectroscopy}},
  \href{https://arxiv.org/abs/2306.17160}{{\ttfamily 2306.17160}}.

\bibitem{Ghoshal:2023fhh}
A.~Ghoshal and A.~Strumia, \emph{{Probing the Dark Matter density with
  gravitational waves from super-massive binary black holes}},
  \href{https://arxiv.org/abs/2306.17158}{{\ttfamily 2306.17158}}.

\bibitem{Deng:2023btv}
H.~Deng, B.~B\'ecsy, X.~Siemens, N.J.~Cornish and D.R.~Madison,
  \emph{{Searching for gravitational wave burst in PTA data with piecewise
  linear functions}},  \href{https://arxiv.org/abs/2306.17130}{{\ttfamily
  2306.17130}}.

\bibitem{Rini:2023eqy}
M.~Rini, \emph{{Researchers Capture Gravitational-Wave Background with Pulsar
  \textquotedblleft{}Antennae\textquotedblright{}}},
  \href{https://doi.org/10.1103/Physics.16.118}{\emph{APS Physics} {\bfseries
  16} (2023) 118}.

\bibitem{Athron:2023mer}
P.~Athron, A.~Fowlie, C.-T.~Lu, L.~Morris, L.~Wu, Z.~Xu et~al., \emph{{Can
  Supercooled Phase Transitions explain the Gravitational Wave Background
  observed by Pulsar Timing Arrays?}},
  \href{https://arxiv.org/abs/2306.17239}{{\ttfamily 2306.17239}}.

\bibitem{Addazi:2023jvg}
A.~Addazi, Y.-F.~Cai, A.~Marciano and L.~Visinelli, \emph{{Have pulsar timing
  array methods detected a cosmological phase transition?}},
  \href{https://arxiv.org/abs/2306.17205}{{\ttfamily 2306.17205}}.

\bibitem{Oikonomou:2023qfz}
V.K.~Oikonomou, \emph{{Flat Energy Spectrum of Primordial Gravitational Waves
  vs Peaks and the NANOGrav 2023 Observation}},
  \href{https://arxiv.org/abs/2306.17351}{{\ttfamily 2306.17351}}.

\bibitem{Kitajima:2023vre}
N.~Kitajima and K.~Nakayama, \emph{{Nanohertz gravitational waves from cosmic
  strings and dark photon dark matter}},
  \href{https://arxiv.org/abs/2306.17390}{{\ttfamily 2306.17390}}.

\bibitem{King:2023cgv}
S.F.~King, D.~Marfatia and M.H.~Rahat, \emph{{Towards distinguishing Dirac from
  Majorana neutrino mass with gravitational waves}},
  \href{https://arxiv.org/abs/2306.05389}{{\ttfamily 2306.05389}}.

\bibitem{Zhao:2022kvz}
Z.-C.~Zhao and S.~Wang, \emph{{Bayesian Implications for the Primordial Black
  Holes from NANOGrav\textquoteright{}s Pulsar-Timing Data Using the
  Scalar-Induced Gravitational Waves}},
  \href{https://doi.org/10.3390/universe9040157}{\emph{Universe} {\bfseries 9}
  (2023) 157} [\href{https://arxiv.org/abs/2211.09450}{{\ttfamily
  2211.09450}}].

\bibitem{Vagnozzi:2020gtf}
S.~Vagnozzi, \emph{{Implications of the NANOGrav results for inflation}},
  \href{https://doi.org/10.1093/mnrasl/slaa203}{\emph{Mon. Not. Roy. Astron.
  Soc.} {\bfseries 502} (2021) L11}
  [\href{https://arxiv.org/abs/2009.13432}{{\ttfamily 2009.13432}}].

\bibitem{Benetti:2021uea}
M.~Benetti, L.L.~Graef and S.~Vagnozzi, \emph{{Primordial gravitational waves
  from NANOGrav: A broken power-law approach}},
  \href{https://doi.org/10.1103/PhysRevD.105.043520}{\emph{Phys. Rev. D}
  {\bfseries 105} (2022) 043520}
  [\href{https://arxiv.org/abs/2111.04758}{{\ttfamily 2111.04758}}].

\bibitem{Chen:2019xse}
Z.-C.~Chen, C.~Yuan and Q.-G.~Huang, \emph{{Pulsar Timing Array Constraints on
  Primordial Black Holes with NANOGrav 11-Year Dataset}},
  \href{https://doi.org/10.1103/PhysRevLett.124.251101}{\emph{Phys. Rev. Lett.}
  {\bfseries 124} (2020) 251101}
  [\href{https://arxiv.org/abs/1910.12239}{{\ttfamily 1910.12239}}].

\bibitem{Kobakhidze:2017mru}
A.~Kobakhidze, C.~Lagger, A.~Manning and J.~Yue, \emph{{Gravitational waves
  from a supercooled electroweak phase transition and their detection with
  pulsar timing arrays}},
  \href{https://doi.org/10.1140/epjc/s10052-017-5132-y}{\emph{Eur. Phys. J. C}
  {\bfseries 77} (2017) 570}
  [\href{https://arxiv.org/abs/1703.06552}{{\ttfamily 1703.06552}}].

\bibitem{Arunasalam:2017ajm}
S.~Arunasalam, A.~Kobakhidze, C.~Lagger, S.~Liang and A.~Zhou, \emph{{Low
  temperature electroweak phase transition in the Standard Model with hidden
  scale invariance}},
  \href{https://doi.org/10.1016/j.physletb.2017.11.017}{\emph{Phys. Lett. B}
  {\bfseries 776} (2018) 48}
  [\href{https://arxiv.org/abs/1709.10322}{{\ttfamily 1709.10322}}].

\bibitem{Niu:2022quw}
X.~Niu, M.H.~Rahat, K.~Srinivasan and W.~Xue, \emph{{Gravitational wave probes
  of massive gauge bosons at the cosmological collider}},
  \href{https://doi.org/10.1088/1475-7516/2023/02/013}{\emph{JCAP} {\bfseries
  02} (2023) 013} [\href{https://arxiv.org/abs/2211.14331}{{\ttfamily
  2211.14331}}].

\bibitem{Niu:2022fki}
X.~Niu, M.H.~Rahat, K.~Srinivasan and W.~Xue, \emph{{Parity-odd and even
  trispectrum from axion inflation}},
  \href{https://doi.org/10.1088/1475-7516/2023/05/018}{\emph{JCAP} {\bfseries
  05} (2023) 018} [\href{https://arxiv.org/abs/2211.14324}{{\ttfamily
  2211.14324}}].

\bibitem{Adshead:2015pva}
P.~Adshead, J.T.~Giblin, T.R.~Scully and E.I.~Sfakianakis,
  \emph{{Gauge-preheating and the end of axion inflation}},
  \href{https://doi.org/10.1088/1475-7516/2015/12/034}{\emph{JCAP} {\bfseries
  12} (2015) 034} [\href{https://arxiv.org/abs/1502.06506}{{\ttfamily
  1502.06506}}].

\bibitem{Adshead:2016iae}
P.~Adshead, J.T.~Giblin, T.R.~Scully and E.I.~Sfakianakis,
  \emph{{Magnetogenesis from axion inflation}},
  \href{https://doi.org/10.1088/1475-7516/2016/10/039}{\emph{JCAP} {\bfseries
  10} (2016) 039} [\href{https://arxiv.org/abs/1606.08474}{{\ttfamily
  1606.08474}}].

\bibitem{Adshead:2018doq}
P.~Adshead, J.T.~Giblin and Z.J.~Weiner, \emph{{Gravitational waves from gauge
  preheating}}, \href{https://doi.org/10.1103/PhysRevD.98.043525}{\emph{Phys.
  Rev. D} {\bfseries 98} (2018) 043525}
  [\href{https://arxiv.org/abs/1805.04550}{{\ttfamily 1805.04550}}].

\bibitem{Adshead:2019lbr}
P.~Adshead, J.T.~Giblin, M.~Pieroni and Z.J.~Weiner, \emph{{Constraining axion
  inflation with gravitational waves from preheating}},
  \href{https://doi.org/10.1103/PhysRevD.101.083534}{\emph{Phys. Rev. D}
  {\bfseries 101} (2020) 083534}
  [\href{https://arxiv.org/abs/1909.12842}{{\ttfamily 1909.12842}}].

\bibitem{Adshead:2019igv}
P.~Adshead, J.T.~Giblin, M.~Pieroni and Z.J.~Weiner, \emph{{Constraining Axion
  Inflation with Gravitational Waves across 29 Decades in Frequency}},
  \href{https://doi.org/10.1103/PhysRevLett.124.171301}{\emph{Phys. Rev. Lett.}
  {\bfseries 124} (2020) 171301}
  [\href{https://arxiv.org/abs/1909.12843}{{\ttfamily 1909.12843}}].

\bibitem{Freese:1990rb}
K.~Freese, J.A.~Frieman and A.V.~Olinto, \emph{{Natural inflation with pseudo -
  Nambu-Goldstone bosons}},
  \href{https://doi.org/10.1103/PhysRevLett.65.3233}{\emph{Phys. Rev. Lett.}
  {\bfseries 65} (1990) 3233}.

\bibitem{Silverstein:2008sg}
E.~Silverstein and A.~Westphal, \emph{{Monodromy in the CMB: Gravity Waves and
  String Inflation}},
  \href{https://doi.org/10.1103/PhysRevD.78.106003}{\emph{Phys. Rev. D}
  {\bfseries 78} (2008) 106003}
  [\href{https://arxiv.org/abs/0803.3085}{{\ttfamily 0803.3085}}].

\bibitem{McAllister:2008hb}
L.~McAllister, E.~Silverstein and A.~Westphal, \emph{{Gravity Waves and Linear
  Inflation from Axion Monodromy}},
  \href{https://doi.org/10.1103/PhysRevD.82.046003}{\emph{Phys. Rev. D}
  {\bfseries 82} (2010) 046003}
  [\href{https://arxiv.org/abs/0808.0706}{{\ttfamily 0808.0706}}].

\bibitem{Kim:2004rp}
J.E.~Kim, H.P.~Nilles and M.~Peloso, \emph{{Completing natural inflation}},
  \href{https://doi.org/10.1088/1475-7516/2005/01/005}{\emph{JCAP} {\bfseries
  01} (2005) 005} [\href{https://arxiv.org/abs/hep-ph/0409138}{{\ttfamily
  hep-ph/0409138}}].

\bibitem{Berg:2009tg}
M.~Berg, E.~Pajer and S.~Sjors, \emph{{Dante's Inferno}},
  \href{https://doi.org/10.1103/PhysRevD.81.103535}{\emph{Phys. Rev. D}
  {\bfseries 81} (2010) 103535}
  [\href{https://arxiv.org/abs/0912.1341}{{\ttfamily 0912.1341}}].

\bibitem{Dimopoulos:2005ac}
S.~Dimopoulos, S.~Kachru, J.~McGreevy and J.G.~Wacker, \emph{{N-flation}},
  \href{https://doi.org/10.1088/1475-7516/2008/08/003}{\emph{JCAP} {\bfseries
  08} (2008) 003} [\href{https://arxiv.org/abs/hep-th/0507205}{{\ttfamily
  hep-th/0507205}}].

\bibitem{Pajer:2013fsa}
E.~Pajer and M.~Peloso, \emph{{A review of Axion Inflation in the era of
  Planck}}, \href{https://doi.org/10.1088/0264-9381/30/21/214002}{\emph{Class.
  Quant. Grav.} {\bfseries 30} (2013) 214002}
  [\href{https://arxiv.org/abs/1305.3557}{{\ttfamily 1305.3557}}].

\bibitem{Anber:2006xt}
M.M.~Anber and L.~Sorbo, \emph{{N-flationary magnetic fields}},
  \href{https://doi.org/10.1088/1475-7516/2006/10/018}{\emph{JCAP} {\bfseries
  10} (2006) 018} [\href{https://arxiv.org/abs/astro-ph/0606534}{{\ttfamily
  astro-ph/0606534}}].

\bibitem{Anber:2009ua}
M.M.~Anber and L.~Sorbo, \emph{{Naturally inflating on steep potentials through
  electromagnetic dissipation}},
  \href{https://doi.org/10.1103/PhysRevD.81.043534}{\emph{Phys. Rev. D}
  {\bfseries 81} (2010) 043534}
  [\href{https://arxiv.org/abs/0908.4089}{{\ttfamily 0908.4089}}].

\bibitem{Cook:2011hg}
J.L.~Cook and L.~Sorbo, \emph{{Particle production during inflation and
  gravitational waves detectable by ground-based interferometers}},
  \href{https://doi.org/10.1103/PhysRevD.85.023534}{\emph{Phys. Rev. D}
  {\bfseries 85} (2012) 023534}
  [\href{https://arxiv.org/abs/1109.0022}{{\ttfamily 1109.0022}}].

\bibitem{Barnaby:2010vf}
N.~Barnaby and M.~Peloso, \emph{{Large Nongaussianity in Axion Inflation}},
  \href{https://doi.org/10.1103/PhysRevLett.106.181301}{\emph{Phys. Rev. Lett.}
  {\bfseries 106} (2011) 181301}
  [\href{https://arxiv.org/abs/1011.1500}{{\ttfamily 1011.1500}}].

\bibitem{Barnaby:2011qe}
N.~Barnaby, E.~Pajer and M.~Peloso, \emph{{Gauge Field Production in Axion
  Inflation: Consequences for Monodromy, non-Gaussianity in the CMB, and
  Gravitational Waves at Interferometers}},
  \href{https://doi.org/10.1103/PhysRevD.85.023525}{\emph{Phys. Rev. D}
  {\bfseries 85} (2012) 023525}
  [\href{https://arxiv.org/abs/1110.3327}{{\ttfamily 1110.3327}}].

\bibitem{Barnaby:2011vw}
N.~Barnaby, R.~Namba and M.~Peloso, \emph{{Phenomenology of a Pseudo-Scalar
  Inflaton: Naturally Large Nongaussianity}},
  \href{https://doi.org/10.1088/1475-7516/2011/04/009}{\emph{JCAP} {\bfseries
  04} (2011) 009} [\href{https://arxiv.org/abs/1102.4333}{{\ttfamily
  1102.4333}}].

\bibitem{Meerburg:2012id}
P.D.~Meerburg and E.~Pajer, \emph{{Observational Constraints on Gauge Field
  Production in Axion Inflation}},
  \href{https://doi.org/10.1088/1475-7516/2013/02/017}{\emph{JCAP} {\bfseries
  02} (2013) 017} [\href{https://arxiv.org/abs/1203.6076}{{\ttfamily
  1203.6076}}].

\bibitem{Anber:2012du}
M.M.~Anber and L.~Sorbo, \emph{{Non-Gaussianities and chiral gravitational
  waves in natural steep inflation}},
  \href{https://doi.org/10.1103/PhysRevD.85.123537}{\emph{Phys. Rev. D}
  {\bfseries 85} (2012) 123537}
  [\href{https://arxiv.org/abs/1203.5849}{{\ttfamily 1203.5849}}].

\bibitem{Linde:2012bt}
A.~Linde, S.~Mooij and E.~Pajer, \emph{{Gauge field production in supergravity
  inflation: Local non-Gaussianity and primordial black holes}},
  \href{https://doi.org/10.1103/PhysRevD.87.103506}{\emph{Phys. Rev. D}
  {\bfseries 87} (2013) 103506}
  [\href{https://arxiv.org/abs/1212.1693}{{\ttfamily 1212.1693}}].

\bibitem{Cheng:2015oqa}
S.-L.~Cheng, W.~Lee and K.-W.~Ng, \emph{{Numerical study of pseudoscalar
  inflation with an axion-gauge field coupling}},
  \href{https://doi.org/10.1103/PhysRevD.93.063510}{\emph{Phys. Rev. D}
  {\bfseries 93} (2016) 063510}
  [\href{https://arxiv.org/abs/1508.00251}{{\ttfamily 1508.00251}}].

\bibitem{Garcia-Bellido:2016dkw}
J.~Garcia-Bellido, M.~Peloso and C.~Unal, \emph{{Gravitational waves at
  interferometer scales and primordial black holes in axion inflation}},
  \href{https://doi.org/10.1088/1475-7516/2016/12/031}{\emph{JCAP} {\bfseries
  12} (2016) 031} [\href{https://arxiv.org/abs/1610.03763}{{\ttfamily
  1610.03763}}].

\bibitem{Domcke:2016bkh}
V.~Domcke, M.~Pieroni and P.~Bin\'etruy, \emph{{Primordial gravitational waves
  for universality classes of pseudoscalar inflation}},
  \href{https://doi.org/10.1088/1475-7516/2016/06/031}{\emph{JCAP} {\bfseries
  06} (2016) 031} [\href{https://arxiv.org/abs/1603.01287}{{\ttfamily
  1603.01287}}].

\bibitem{Domcke:2016mbx}
V.~Domcke, \emph{{Probing inflation models with gravitational waves}},  in
  \emph{{51st Rencontres de Moriond on Cosmology}}, pp.~205--208, 5, 2016
  [\href{https://arxiv.org/abs/1605.06364}{{\ttfamily 1605.06364}}].

\bibitem{Peloso:2016gqs}
M.~Peloso, L.~Sorbo and C.~Unal, \emph{{Rolling axions during inflation:
  perturbativity and signatures}},
  \href{https://doi.org/10.1088/1475-7516/2016/09/001}{\emph{JCAP} {\bfseries
  09} (2016) 001} [\href{https://arxiv.org/abs/1606.00459}{{\ttfamily
  1606.00459}}].

\bibitem{Domcke:2018eki}
V.~Domcke and K.~Mukaida, \emph{{Gauge Field and Fermion Production during
  Axion Inflation}},
  \href{https://doi.org/10.1088/1475-7516/2018/11/020}{\emph{JCAP} {\bfseries
  11} (2018) 020} [\href{https://arxiv.org/abs/1806.08769}{{\ttfamily
  1806.08769}}].

\bibitem{Cuissa:2018oiw}
J.R.C.~Cuissa and D.G.~Figueroa, \emph{{Lattice formulation of axion inflation.
  Application to preheating}},
  \href{https://doi.org/10.1088/1475-7516/2019/06/002}{\emph{JCAP} {\bfseries
  06} (2019) 002} [\href{https://arxiv.org/abs/1812.03132}{{\ttfamily
  1812.03132}}].

\bibitem{KAGRA:2021kbb}
{\scshape KAGRA, Virgo, LIGO Scientific} collaboration, \emph{{Upper limits on
  the isotropic gravitational-wave background from Advanced LIGO and Advanced
  Virgo\textquoteright{}s third observing run}},
  \href{https://doi.org/10.1103/PhysRevD.104.022004}{\emph{Phys. Rev. D}
  {\bfseries 104} (2021) 022004}
  [\href{https://arxiv.org/abs/2101.12130}{{\ttfamily 2101.12130}}].

\bibitem{Carrasco:2015rva}
J.J.M.~Carrasco, R.~Kallosh and A.~Linde, \emph{{Cosmological Attractors and
  Initial Conditions for Inflation}},
  \href{https://doi.org/10.1103/PhysRevD.92.063519}{\emph{Phys. Rev. D}
  {\bfseries 92} (2015) 063519}
  [\href{https://arxiv.org/abs/1506.00936}{{\ttfamily 1506.00936}}].

\bibitem{Planck:2018jri}
{\scshape Planck} collaboration, \emph{{Planck 2018 results. X. Constraints on
  inflation}}, \href{https://doi.org/10.1051/0004-6361/201833887}{\emph{Astron.
  Astrophys.} {\bfseries 641} (2020) A10}
  [\href{https://arxiv.org/abs/1807.06211}{{\ttfamily 1807.06211}}].

\bibitem{Orlando:2020oko}
G.~Orlando, M.~Pieroni and A.~Ricciardone, \emph{{Measuring Parity Violation in
  the Stochastic Gravitational Wave Background with the LISA-Taiji network}},
  \href{https://doi.org/10.1088/1475-7516/2021/03/069}{\emph{JCAP} {\bfseries
  03} (2021) 069} [\href{https://arxiv.org/abs/2011.07059}{{\ttfamily
  2011.07059}}].

\bibitem{Badger:2021enh}
C.~Badger and M.~Sakellariadou, \emph{{Detecting parity violation from axion
  inflation with third generation detectors}},
  \href{https://arxiv.org/abs/2112.04650}{{\ttfamily 2112.04650}}.

\bibitem{Chen:2009zp}
X.~Chen and Y.~Wang, \emph{{Quasi-Single Field Inflation and
  Non-Gaussianities}},
  \href{https://doi.org/10.1088/1475-7516/2010/04/027}{\emph{JCAP} {\bfseries
  04} (2010) 027} [\href{https://arxiv.org/abs/0911.3380}{{\ttfamily
  0911.3380}}].

\bibitem{Chen:2009we}
X.~Chen and Y.~Wang, \emph{{Large non-Gaussianities with Intermediate Shapes
  from Quasi-Single Field Inflation}},
  \href{https://doi.org/10.1103/PhysRevD.81.063511}{\emph{Phys. Rev. D}
  {\bfseries 81} (2010) 063511}
  [\href{https://arxiv.org/abs/0909.0496}{{\ttfamily 0909.0496}}].

\bibitem{Baumann:2011nk}
D.~Baumann and D.~Green, \emph{{Signatures of Supersymmetry from the Early
  Universe}}, \href{https://doi.org/10.1103/PhysRevD.85.103520}{\emph{Phys.
  Rev. D} {\bfseries 85} (2012) 103520}
  [\href{https://arxiv.org/abs/1109.0292}{{\ttfamily 1109.0292}}].

\bibitem{Arkani-Hamed:2015bza}
N.~Arkani-Hamed and J.~Maldacena, \emph{{Cosmological Collider Physics}},
  \href{https://arxiv.org/abs/1503.08043}{{\ttfamily 1503.08043}}.

\bibitem{Chen:2016nrs}
X.~Chen, Y.~Wang and Z.-Z.~Xianyu, \emph{{Loop Corrections to Standard Model
  Fields in Inflation}},
  \href{https://doi.org/10.1007/JHEP08(2016)051}{\emph{JHEP} {\bfseries 08}
  (2016) 051} [\href{https://arxiv.org/abs/1604.07841}{{\ttfamily
  1604.07841}}].

\bibitem{Lee:2016vti}
H.~Lee, D.~Baumann and G.L.~Pimentel, \emph{{Non-Gaussianity as a Particle
  Detector}}, \href{https://doi.org/10.1007/JHEP12(2016)040}{\emph{JHEP}
  {\bfseries 12} (2016) 040}
  [\href{https://arxiv.org/abs/1607.03735}{{\ttfamily 1607.03735}}].

\bibitem{Meerburg:2016zdz}
P.D.~Meerburg, M.~M\"unchmeyer, J.B.~Mu\~noz and X.~Chen, \emph{{Prospects for
  Cosmological Collider Physics}},
  \href{https://doi.org/10.1088/1475-7516/2017/03/050}{\emph{JCAP} {\bfseries
  03} (2017) 050} [\href{https://arxiv.org/abs/1610.06559}{{\ttfamily
  1610.06559}}].

\bibitem{Chen:2016uwp}
X.~Chen, Y.~Wang and Z.-Z.~Xianyu, \emph{{Standard Model Background of the
  Cosmological Collider}},
  \href{https://doi.org/10.1103/PhysRevLett.118.261302}{\emph{Phys. Rev. Lett.}
  {\bfseries 118} (2017) 261302}
  [\href{https://arxiv.org/abs/1610.06597}{{\ttfamily 1610.06597}}].

\bibitem{Chen:2016hrz}
X.~Chen, Y.~Wang and Z.-Z.~Xianyu, \emph{{Standard Model Mass Spectrum in
  Inflationary Universe}},
  \href{https://doi.org/10.1007/JHEP04(2017)058}{\emph{JHEP} {\bfseries 04}
  (2017) 058} [\href{https://arxiv.org/abs/1612.08122}{{\ttfamily
  1612.08122}}].

\bibitem{An:2017hlx}
H.~An, M.~McAneny, A.K.~Ridgway and M.B.~Wise, \emph{{Quasi Single Field
  Inflation in the non-perturbative regime}},
  \href{https://doi.org/10.1007/JHEP06(2018)105}{\emph{JHEP} {\bfseries 06}
  (2018) 105} [\href{https://arxiv.org/abs/1706.09971}{{\ttfamily
  1706.09971}}].

\bibitem{Kumar:2017ecc}
S.~Kumar and R.~Sundrum, \emph{{Heavy-Lifting of Gauge Theories By Cosmic
  Inflation}}, \href{https://doi.org/10.1007/JHEP05(2018)011}{\emph{JHEP}
  {\bfseries 05} (2018) 011}
  [\href{https://arxiv.org/abs/1711.03988}{{\ttfamily 1711.03988}}].

\bibitem{Chen:2018xck}
X.~Chen, Y.~Wang and Z.-Z.~Xianyu, \emph{{Neutrino Signatures in Primordial
  Non-Gaussianities}},
  \href{https://doi.org/10.1007/JHEP09(2018)022}{\emph{JHEP} {\bfseries 09}
  (2018) 022} [\href{https://arxiv.org/abs/1805.02656}{{\ttfamily
  1805.02656}}].

\bibitem{Wu:2018lmx}
Y.-P.~Wu, \emph{{Higgs as heavy-lifted physics during inflation}},
  \href{https://doi.org/10.1007/JHEP04(2019)125}{\emph{JHEP} {\bfseries 04}
  (2019) 125} [\href{https://arxiv.org/abs/1812.10654}{{\ttfamily
  1812.10654}}].

\bibitem{Li:2019ves}
L.~Li, T.~Nakama, C.M.~Sou, Y.~Wang and S.~Zhou, \emph{{Gravitational
  Production of Superheavy Dark Matter and Associated Cosmological
  Signatures}}, \href{https://doi.org/10.1007/JHEP07(2019)067}{\emph{JHEP}
  {\bfseries 07} (2019) 067}
  [\href{https://arxiv.org/abs/1903.08842}{{\ttfamily 1903.08842}}].

\bibitem{Lu:2019tjj}
S.~Lu, Y.~Wang and Z.-Z.~Xianyu, \emph{{A Cosmological Higgs Collider}},
  \href{https://doi.org/10.1007/JHEP02(2020)011}{\emph{JHEP} {\bfseries 02}
  (2020) 011} [\href{https://arxiv.org/abs/1907.07390}{{\ttfamily
  1907.07390}}].

\bibitem{Hook:2019zxa}
A.~Hook, J.~Huang and D.~Racco, \emph{{Searches for other vacua. Part II. A new
  Higgstory at the cosmological collider}},
  \href{https://doi.org/10.1007/JHEP01(2020)105}{\emph{JHEP} {\bfseries 01}
  (2020) 105} [\href{https://arxiv.org/abs/1907.10624}{{\ttfamily
  1907.10624}}].

\bibitem{Hook:2019vcn}
A.~Hook, J.~Huang and D.~Racco, \emph{{Minimal signatures of the Standard Model
  in non-Gaussianities}},
  \href{https://doi.org/10.1103/PhysRevD.101.023519}{\emph{Phys. Rev. D}
  {\bfseries 101} (2020) 023519}
  [\href{https://arxiv.org/abs/1908.00019}{{\ttfamily 1908.00019}}].

\bibitem{Kumar:2019ebj}
S.~Kumar and R.~Sundrum, \emph{{Cosmological Collider Physics and the
  Curvaton}}, \href{https://doi.org/10.1007/JHEP04(2020)077}{\emph{JHEP}
  {\bfseries 04} (2020) 077}
  [\href{https://arxiv.org/abs/1908.11378}{{\ttfamily 1908.11378}}].

\bibitem{Wang:2019gbi}
L.-T.~Wang and Z.-Z.~Xianyu, \emph{{In Search of Large Signals at the
  Cosmological Collider}},
  \href{https://doi.org/10.1007/JHEP02(2020)044}{\emph{JHEP} {\bfseries 02}
  (2020) 044} [\href{https://arxiv.org/abs/1910.12876}{{\ttfamily
  1910.12876}}].

\bibitem{Wang:2020uic}
Y.~Wang and Y.~Zhu, \emph{{Cosmological Collider Signatures of Massive Vectors
  from Non-Gaussian Gravitational Waves}},
  \href{https://doi.org/10.1088/1475-7516/2020/04/049}{\emph{JCAP} {\bfseries
  04} (2020) 049} [\href{https://arxiv.org/abs/2001.03879}{{\ttfamily
  2001.03879}}].

\bibitem{Wang:2020ioa}
L.-T.~Wang and Z.-Z.~Xianyu, \emph{{Gauge Boson Signals at the Cosmological
  Collider}}, \href{https://doi.org/10.1007/JHEP11(2020)082}{\emph{JHEP}
  {\bfseries 11} (2020) 082}
  [\href{https://arxiv.org/abs/2004.02887}{{\ttfamily 2004.02887}}].

\bibitem{Aoki:2020zbj}
S.~Aoki and M.~Yamaguchi, \emph{{Disentangling mass spectra of multiple fields
  in cosmological collider}},
  \href{https://doi.org/10.1007/JHEP04(2021)127}{\emph{JHEP} {\bfseries 04}
  (2021) 127} [\href{https://arxiv.org/abs/2012.13667}{{\ttfamily
  2012.13667}}].

\bibitem{Maru:2021ezc}
N.~Maru and A.~Okawa, \emph{{Non-Gaussianity from $X, Y$ gauge bosons in
  Cosmological Collider Physics}},
  \href{https://arxiv.org/abs/2101.10634}{{\ttfamily 2101.10634}}.

\bibitem{Lu:2021gso}
S.~Lu, \emph{{Axion isocurvature collider}},
  \href{https://doi.org/10.1007/JHEP04(2022)157}{\emph{JHEP} {\bfseries 04}
  (2022) 157} [\href{https://arxiv.org/abs/2103.05958}{{\ttfamily
  2103.05958}}].

\bibitem{Wang:2021qez}
L.-T.~Wang, Z.-Z.~Xianyu and Y.-M.~Zhong, \emph{{Precision calculation of
  inflation correlators at one loop}},
  \href{https://doi.org/10.1007/JHEP02(2022)085}{\emph{JHEP} {\bfseries 02}
  (2022) 085} [\href{https://arxiv.org/abs/2109.14635}{{\ttfamily
  2109.14635}}].

\bibitem{Tong:2021wai}
X.~Tong, Y.~Wang and Y.~Zhu, \emph{{Cutting rule for cosmological collider
  signals: a bulk evolution perspective}},
  \href{https://doi.org/10.1007/JHEP03(2022)181}{\emph{JHEP} {\bfseries 03}
  (2022) 181} [\href{https://arxiv.org/abs/2112.03448}{{\ttfamily
  2112.03448}}].

\bibitem{Cui:2021iie}
Y.~Cui and Z.-Z.~Xianyu, \emph{{Probing Leptogenesis with the Cosmological
  Collider}}, \href{https://doi.org/10.1103/PhysRevLett.129.111301}{\emph{Phys.
  Rev. Lett.} {\bfseries 129} (2022) 111301}
  [\href{https://arxiv.org/abs/2112.10793}{{\ttfamily 2112.10793}}].

\bibitem{Pinol:2021aun}
L.~Pinol, S.~Aoki, S.~Renaux-Petel and M.~Yamaguchi, \emph{{Inflationary flavor
  oscillations and the cosmic spectroscopy}},
  \href{https://arxiv.org/abs/2112.05710}{{\ttfamily 2112.05710}}.

\bibitem{Tong:2022cdz}
X.~Tong and Z.-Z.~Xianyu, \emph{{Large spin-2 signals at the cosmological
  collider}}, \href{https://doi.org/10.1007/JHEP10(2022)194}{\emph{JHEP}
  {\bfseries 10} (2022) 194}
  [\href{https://arxiv.org/abs/2203.06349}{{\ttfamily 2203.06349}}].

\bibitem{Reece:2022soh}
M.~Reece, L.-T.~Wang and Z.-Z.~Xianyu, \emph{{Large-Field Inflation and the
  Cosmological Collider}},  \href{https://arxiv.org/abs/2204.11869}{{\ttfamily
  2204.11869}}.

\bibitem{Jazayeri:2022kjy}
S.~Jazayeri and S.~Renaux-Petel, \emph{{Cosmological Bootstrap in Slow
  Motion}},  \href{https://arxiv.org/abs/2205.10340}{{\ttfamily 2205.10340}}.

\bibitem{Pimentel:2022fsc}
G.L.~Pimentel and D.-G.~Wang, \emph{{Boostless cosmological collider
  bootstrap}}, \href{https://doi.org/10.1007/JHEP10(2022)177}{\emph{JHEP}
  {\bfseries 10} (2022) 177}
  [\href{https://arxiv.org/abs/2205.00013}{{\ttfamily 2205.00013}}].

\bibitem{Chen:2022vzh}
X.~Chen, R.~Ebadi and S.~Kumar, \emph{{Classical cosmological collider physics
  and primordial features}},
  \href{https://doi.org/10.1088/1475-7516/2022/08/083}{\emph{JCAP} {\bfseries
  08} (2022) 083} [\href{https://arxiv.org/abs/2205.01107}{{\ttfamily
  2205.01107}}].

\bibitem{Qin:2022lva}
Z.~Qin and Z.-Z.~Xianyu, \emph{{Phase information in cosmological collider
  signals}}, \href{https://doi.org/10.1007/JHEP10(2022)192}{\emph{JHEP}
  {\bfseries 10} (2022) 192}
  [\href{https://arxiv.org/abs/2205.01692}{{\ttfamily 2205.01692}}].

\bibitem{Maru:2022bhr}
N.~Maru and A.~Okawa, \emph{{Cosmological Collider Signals of Non-Gaussianity
  from Higgs boson in GUT}},
  \href{https://arxiv.org/abs/2206.06651}{{\ttfamily 2206.06651}}.

\bibitem{Aoki:2023tjm}
S.~Aoki, \emph{{Continuous spectrum on cosmological collider}},
  \href{https://doi.org/10.1088/1475-7516/2023/04/002}{\emph{JCAP} {\bfseries
  04} (2023) 002} [\href{https://arxiv.org/abs/2301.07920}{{\ttfamily
  2301.07920}}].

\bibitem{Chen:2023txq}
X.~Chen, J.~Fan and L.~Li, \emph{{New inflationary probes of axion dark
  matter}},  \href{https://arxiv.org/abs/2303.03406}{{\ttfamily 2303.03406}}.

\bibitem{Graham:2015rva}
P.W.~Graham, J.~Mardon and S.~Rajendran, \emph{{Vector Dark Matter from
  Inflationary Fluctuations}},
  \href{https://doi.org/10.1103/PhysRevD.93.103520}{\emph{Phys. Rev. D}
  {\bfseries 93} (2016) 103520}
  [\href{https://arxiv.org/abs/1504.02102}{{\ttfamily 1504.02102}}].

\bibitem{Ema:2019yrd}
Y.~Ema, K.~Nakayama and Y.~Tang, \emph{{Production of purely gravitational dark
  matter: the case of fermion and vector boson}},
  \href{https://doi.org/10.1007/JHEP07(2019)060}{\emph{JHEP} {\bfseries 07}
  (2019) 060} [\href{https://arxiv.org/abs/1903.10973}{{\ttfamily
  1903.10973}}].

\bibitem{Ahmed:2020fhc}
A.~Ahmed, B.~Grzadkowski and A.~Socha, \emph{{Gravitational production of
  vector dark matter}},
  \href{https://doi.org/10.1007/JHEP08(2020)059}{\emph{JHEP} {\bfseries 08}
  (2020) 059} [\href{https://arxiv.org/abs/2005.01766}{{\ttfamily
  2005.01766}}].

\bibitem{Kolb:2020fwh}
E.W.~Kolb and A.J.~Long, \emph{{Completely dark photons from gravitational
  particle production during the inflationary era}},
  \href{https://doi.org/10.1007/JHEP03(2021)283}{\emph{JHEP} {\bfseries 03}
  (2021) 283} [\href{https://arxiv.org/abs/2009.03828}{{\ttfamily
  2009.03828}}].

\bibitem{Fong:2022cmq}
C.S.~Fong, M.H.~Rahat and S.~Saad, \emph{{BBN photodisintegration constraints
  on gravitationally produced vector bosons}},
  \href{https://doi.org/10.1007/JHEP11(2022)067}{\emph{JHEP} {\bfseries 11}
  (2022) 067} [\href{https://arxiv.org/abs/2206.02802}{{\ttfamily
  2206.02802}}].

\bibitem{Weinberg:2008zzc}
S.~Weinberg, \emph{{Cosmology}} (2008).

\bibitem{Weinberg:2005vy}
S.~Weinberg, \emph{{Quantum contributions to cosmological correlations}},
  \href{https://doi.org/10.1103/PhysRevD.72.043514}{\emph{Phys. Rev. D}
  {\bfseries 72} (2005) 043514}
  [\href{https://arxiv.org/abs/hep-th/0506236}{{\ttfamily hep-th/0506236}}].

\bibitem{WMAP:2010qai}
{\scshape WMAP} collaboration, \emph{{Seven-Year Wilkinson Microwave Anisotropy
  Probe (WMAP) Observations: Cosmological Interpretation}},
  \href{https://doi.org/10.1088/0067-0049/192/2/18}{\emph{Astrophys. J. Suppl.}
  {\bfseries 192} (2011) 18} [\href{https://arxiv.org/abs/1001.4538}{{\ttfamily
  1001.4538}}].

\bibitem{Bunn:1996py}
E.F.~Bunn, A.R.~Liddle and M.J.~White, \emph{{Four-year COBE normalization of
  inflationary cosmologies}},
  \href{https://doi.org/10.1103/PhysRevD.54.R5917}{\emph{Phys. Rev. D}
  {\bfseries 54} (1996) R5917}
  [\href{https://arxiv.org/abs/astro-ph/9607038}{{\ttfamily
  astro-ph/9607038}}].

\bibitem{Planck:2019kim}
{\scshape Planck} collaboration, \emph{{Planck 2018 results. IX. Constraints on
  primordial non-Gaussianity}},
  \href{https://doi.org/10.1051/0004-6361/201935891}{\emph{Astron. Astrophys.}
  {\bfseries 641} (2020) A9}
  [\href{https://arxiv.org/abs/1905.05697}{{\ttfamily 1905.05697}}].

\bibitem{Namikawa:2019tax}
T.~Namikawa, S.~Saga, D.~Yamauchi and A.~Taruya, \emph{{CMB Constraints on the
  Stochastic Gravitational-Wave Background at Mpc scales}},
  \href{https://doi.org/10.1103/PhysRevD.100.021303}{\emph{Phys. Rev. D}
  {\bfseries 100} (2019) 021303}
  [\href{https://arxiv.org/abs/1904.02115}{{\ttfamily 1904.02115}}].

\bibitem{LIGOScientific:2022sts}
{\scshape LIGO Scientific, VIRGO, KAGRA} collaboration, \emph{{Search for
  gravitational-wave transients associated with magnetar bursts in Advanced
  LIGO and Advanced Virgo data from the third observing run}},
  \href{https://arxiv.org/abs/2210.10931}{{\ttfamily 2210.10931}}.

\bibitem{Jiang:2022uxp}
Y.~Jiang and Q.-G.~Huang, \emph{{Upper limits on the polarized isotropic
  stochastic gravitational-wave background from advanced LIGO-Virgo's first
  three observing runs}},
  \href{https://doi.org/10.1088/1475-7516/2023/02/026}{\emph{JCAP} {\bfseries
  02} (2023) 026} [\href{https://arxiv.org/abs/2210.09952}{{\ttfamily
  2210.09952}}].

\bibitem{Galante:2014ifa}
M.~Galante, R.~Kallosh, A.~Linde and D.~Roest, \emph{{Unity of Cosmological
  Inflation Attractors}},
  \href{https://doi.org/10.1103/PhysRevLett.114.141302}{\emph{Phys. Rev. Lett.}
  {\bfseries 114} (2015) 141302}
  [\href{https://arxiv.org/abs/1412.3797}{{\ttfamily 1412.3797}}].

\bibitem{Kallosh:2013maa}
R.~Kallosh and A.~Linde, \emph{{Non-minimal Inflationary Attractors}},
  \href{https://doi.org/10.1088/1475-7516/2013/10/033}{\emph{JCAP} {\bfseries
  10} (2013) 033} [\href{https://arxiv.org/abs/1307.7938}{{\ttfamily
  1307.7938}}].

\bibitem{Kallosh:2013pby}
R.~Kallosh and A.~Linde, \emph{{Superconformal generalization of the chaotic
  inflation model $\frac{\lambda}{4} \phi^{4} - \frac{\xi}{2} \phi^{2}R$}},
  \href{https://doi.org/10.1088/1475-7516/2013/06/027}{\emph{JCAP} {\bfseries
  06} (2013) 027} [\href{https://arxiv.org/abs/1306.3211}{{\ttfamily
  1306.3211}}].

\bibitem{Kallosh:2013yoa}
R.~Kallosh, A.~Linde and D.~Roest, \emph{{Superconformal Inflationary
  $\alpha$-Attractors}},
  \href{https://doi.org/10.1007/JHEP11(2013)198}{\emph{JHEP} {\bfseries 11}
  (2013) 198} [\href{https://arxiv.org/abs/1311.0472}{{\ttfamily 1311.0472}}].

\bibitem{Kallosh:2013hoa}
R.~Kallosh and A.~Linde, \emph{{Universality Class in Conformal Inflation}},
  \href{https://doi.org/10.1088/1475-7516/2013/07/002}{\emph{JCAP} {\bfseries
  07} (2013) 002} [\href{https://arxiv.org/abs/1306.5220}{{\ttfamily
  1306.5220}}].

\bibitem{Kallosh:2013lkr}
R.~Kallosh and A.~Linde, \emph{{Superconformal generalizations of the
  Starobinsky model}},
  \href{https://doi.org/10.1088/1475-7516/2013/06/028}{\emph{JCAP} {\bfseries
  06} (2013) 028} [\href{https://arxiv.org/abs/1306.3214}{{\ttfamily
  1306.3214}}].

\bibitem{Mitridate:2023oar}
A.~Mitridate, D.~Wright, R.~von Eckardstein, T.~Schr\"oder, J.~Nay, K.~Olum
  et~al., \emph{{PTArcade}},
  \href{https://arxiv.org/abs/2306.16377}{{\ttfamily 2306.16377}}.

\bibitem{Janssen:2014dka}
G.~Janssen et~al., \emph{{Gravitational wave astronomy with the SKA}},
  \href{https://doi.org/10.22323/1.215.0037}{\emph{PoS} {\bfseries AASKA14}
  (2015) 037} [\href{https://arxiv.org/abs/1501.00127}{{\ttfamily
  1501.00127}}].

\bibitem{Sesana:2019vho}
A.~Sesana et~al., \emph{{Unveiling the gravitational universe at $\mu$-Hz
  frequencies}}, \href{https://doi.org/10.1007/s10686-021-09709-9}{\emph{Exper.
  Astron.} {\bfseries 51} (2021) 1333}
  [\href{https://arxiv.org/abs/1908.11391}{{\ttfamily 1908.11391}}].

\bibitem{LISA:2017pwj}
{\scshape LISA} collaboration, \emph{{Laser Interferometer Space Antenna}},
  \href{https://arxiv.org/abs/1702.00786}{{\ttfamily 1702.00786}}.

\bibitem{Kudoh:2005as}
H.~Kudoh, A.~Taruya, T.~Hiramatsu and Y.~Himemoto, \emph{{Detecting a
  gravitational-wave background with next-generation space interferometers}},
  \href{https://doi.org/10.1103/PhysRevD.73.064006}{\emph{Phys. Rev. D}
  {\bfseries 73} (2006) 064006}
  [\href{https://arxiv.org/abs/gr-qc/0511145}{{\ttfamily gr-qc/0511145}}].

\bibitem{Kawamura:2020pcg}
S.~Kawamura et~al., \emph{{Current status of space gravitational wave antenna
  DECIGO and B-DECIGO}},
  \href{https://doi.org/10.1093/ptep/ptab019}{\emph{PTEP} {\bfseries 2021}
  (2021) 05A105} [\href{https://arxiv.org/abs/2006.13545}{{\ttfamily
  2006.13545}}].

\bibitem{Harry:2006fi}
G.M.~Harry, P.~Fritschel, D.A.~Shaddock, W.~Folkner and E.S.~Phinney,
  \emph{{Laser interferometry for the big bang observer}},
  \href{https://doi.org/10.1088/0264-9381/23/15/008}{\emph{Class. Quant. Grav.}
  {\bfseries 23} (2006) 4887}.

\bibitem{AEDGE:2019nxb}
{\scshape AEDGE} collaboration, \emph{{AEDGE: Atomic Experiment for Dark Matter
  and Gravity Exploration in Space}},
  \href{https://doi.org/10.1140/epjqt/s40507-020-0080-0}{\emph{EPJ Quant.
  Technol.} {\bfseries 7} (2020) 6}
  [\href{https://arxiv.org/abs/1908.00802}{{\ttfamily 1908.00802}}].

\bibitem{Badurina:2019hst}
L.~Badurina et~al., \emph{{AION: An Atom Interferometer Observatory and
  Network}}, \href{https://doi.org/10.1088/1475-7516/2020/05/011}{\emph{JCAP}
  {\bfseries 05} (2020) 011}
  [\href{https://arxiv.org/abs/1911.11755}{{\ttfamily 1911.11755}}].

\bibitem{LIGOScientific:2016wof}
{\scshape LIGO Scientific} collaboration, \emph{{Exploring the Sensitivity of
  Next Generation Gravitational Wave Detectors}},
  \href{https://doi.org/10.1088/1361-6382/aa51f4}{\emph{Class. Quant. Grav.}
  {\bfseries 34} (2017) 044001}
  [\href{https://arxiv.org/abs/1607.08697}{{\ttfamily 1607.08697}}].

\bibitem{Hild:2008ng}
S.~Hild, S.~Chelkowski and A.~Freise, \emph{{Pushing towards the ET sensitivity
  using 'conventional' technology}},
  \href{https://arxiv.org/abs/0810.0604}{{\ttfamily 0810.0604}}.

\bibitem{Planck:2018vyg}
{\scshape Planck} collaboration, \emph{{Planck 2018 results. VI. Cosmological
  parameters}},
  \href{https://doi.org/10.1051/0004-6361/201833910}{\emph{Astron. Astrophys.}
  {\bfseries 641} (2020) A6}
  [\href{https://arxiv.org/abs/1807.06209}{{\ttfamily 1807.06209}}].

\bibitem{Josan:2009qn}
A.S.~Josan, A.M.~Green and K.A.~Malik, \emph{{Generalised constraints on the
  curvature perturbation from primordial black holes}},
  \href{https://doi.org/10.1103/PhysRevD.79.103520}{\emph{Phys. Rev. D}
  {\bfseries 79} (2009) 103520}
  [\href{https://arxiv.org/abs/0903.3184}{{\ttfamily 0903.3184}}].

\bibitem{Carr:2020gox}
B.~Carr, K.~Kohri, Y.~Sendouda and J.~Yokoyama, \emph{{Constraints on
  primordial black holes}},
  \href{https://doi.org/10.1088/1361-6633/ac1e31}{\emph{Rept. Prog. Phys.}
  {\bfseries 84} (2021) 116902}
  [\href{https://arxiv.org/abs/2002.12778}{{\ttfamily 2002.12778}}].

\bibitem{Caravano:2022epk}
A.~Caravano, E.~Komatsu, K.D.~Lozanov and J.~Weller, \emph{{Lattice Simulations
  of Axion-U(1) Inflation}},
  \href{https://arxiv.org/abs/2204.12874}{{\ttfamily 2204.12874}}.

\bibitem{Baumann:2007zm}
D.~Baumann, P.J.~Steinhardt, K.~Takahashi and K.~Ichiki, \emph{{Gravitational
  Wave Spectrum Induced by Primordial Scalar Perturbations}},
  \href{https://doi.org/10.1103/PhysRevD.76.084019}{\emph{Phys. Rev. D}
  {\bfseries 76} (2007) 084019}
  [\href{https://arxiv.org/abs/hep-th/0703290}{{\ttfamily hep-th/0703290}}].

\end{thebibliography}\endgroup
\newpage
\bibliographystyle{JHEP}
\end{document}